\documentclass[11pt]{article}
\usepackage{amsmath,amssymb,color,epsfig,cite}
\usepackage{graphicx}
\usepackage{subfigure}
\usepackage{setspace}

\usepackage{amsfonts}

\newcommand{\hoch}[1]{$\, ^{#1}$}

\makeatletter
\@addtoreset{equation}{section}
\makeatother

\newcommand{\be}{\begin{equation}}
\newcommand{\ee}{\end{equation}}
\newcommand{\bea}{\setlength\arraycolsep{2pt} \begin{eqnarray}}
\newcommand{\eea}{\end{eqnarray}}
\newcommand{\nn}{\nonumber}

\def\ft#1#2{{\textstyle{\frac{\scriptstyle #1}{\scriptstyle #2} } }}
\def\fft#1#2{{\frac{#1}{#2}}}

\def\del{{\partial}}

\newcommand{\bpm}{\begin{pmatrix}}
\newcommand{\epm}{\end{pmatrix}}

\def\ft#1#2{{\textstyle{\frac{\scriptstyle #1}{\scriptstyle #2} } }}
\def\fft#1#2{{\frac{#1}{#2}}}

\def\0{{\sst{(0)}}}
\def\1{{\sst{(1)}}}
\def\2{{\sst{(2)}}}
\def\3{{\sst{(3)}}}
\def\4{{\sst{(4)}}}
\def\5{{\sst{(5)}}}
\def\6{{\sst{(6)}}}
\def\7{{\sst{(7)}}}
\def\8{{\sst{(8)}}}

\def\sst#1{{\scriptscriptstyle #1}}
\def\oneone{\rlap 1\mkern4mu{\rm l}}

\def\del{{\partial}}

\def\cA{{{\cal A}}}
\def\cF{{{\cal F}}}

\def\omicron{{o}}
\def\Gj{{G\!j}}

\usepackage{mathrsfs}
\def\scri{\mathscr{I}}

\usepackage{enumitem} 
\usepackage{xcolor}
\usepackage{hyperref}
\hypersetup{hidelinks}

\renewcommand{\O}{\mathcal{O}}

\begin{document}

\begin{flushright}
\hfill {MI-HET-760
\vspace{-1mm}

QMUL-PH-21-37}
\end{flushright}

\vspace{15pt}
\begin{center}
{\Large {\bf Asymptotic Weyl Double Copy}}

\vspace{15pt}

\vspace{15pt}
{\bf Hadi Godazgar\hoch{1}, Mahdi Godazgar\hoch{2}, Ricardo Monteiro\hoch{3}, 

David Peinador Veiga\hoch{3} and C.N. Pope\hoch{4,5}}

\vspace{10pt}

\hoch{1} {\it Max-Planck-Institut f\"ur Gravitationsphysik (Albert-Einstein-Institut), \\
M\"uhlenberg 1, D-14476 Potsdam, Germany.}

\vspace{5pt}

\hoch{2} {\it School of Mathematical Sciences,
Queen Mary University of London, \\
Mile End Road, E1 4NS, UK.}

\vspace{5pt}

\hoch{3} {\it Centre for Theoretical Physics, Department of Physics and Astronomy, \\
Queen Mary University of London, Mile End Road, E1 4NS, UK.}

\vspace{5pt}

\hoch{4} {\it George P. \& Cynthia Woods Mitchell  Institute
for Fundamental Physics and Astronomy,\\
Texas A\&M University, College Station, TX 77843, USA.}

\vspace{5pt}

\hoch{5}{\it DAMTP, Centre for Mathematical Sciences,\\
 Cambridge University, Wilberforce Road, Cambridge CB3 OWA, UK.}

 \vspace{15pt}
 
\today

\vspace{20pt}

\underline{ABSTRACT}
\end{center}

\noindent A characteristic value formulation of the Weyl double copy leads to an asymptotic formulation. We find that the Weyl double copy holds asymptotically in cases where the full solution is algebraically general, using rotating STU supergravity black holes as an example. The asymptotic formulation provides clues regarding the relation between asymptotic symmetries that follows from the double copy.  Using the C-metric as an example, we show that a previous interpretation of this gravity solution as a superrotation has a single copy analogue relating the appropriate Li\'enard-Wiechert potential to a large gauge transformation.

\thispagestyle{empty}

\vfill
\noindent Emails:\hspace{-1mm} hadi.godazgar@aei.mpg.de, m.godazgar@qmul.ac.uk, ricardo.monteiro@qmul.ac.uk, d.peinadorveiga@qmul.ac.uk, pope@physics.tamu.edu

\pagebreak

\tableofcontents

\section{Introduction}

The formulation of gravity as a `double copy' of gauge theory has been remarkably fruitful over the past decade or so; see \cite{Bern:2019prr} for a review. It originated in the realisation of gravity and gauge theory as low-energy limits of closed and open strings, due to the double copy relation between the respective scattering amplitudes \cite{Kawai:1985xq}. Particularly after the framework of \cite{DC,DC1}, this general idea has been exploited to great effect to study a range of perturbative problems, from the ultraviolet divergences of supergravity theories to the classical dynamics of black hole binaries.

In this paper, we will focus on one strand of this progress, which is concerned with the double copy interpretation of classical gravity solutions. Since the double copy had been understood as a property of perturbative gravity, it came as a surprise that exact black hole spacetimes  -- e.g.~the Kerr solution -- could admit a straightforward interpretation as a double copy of a gauge theory solution \cite{Monteiro:2014cda,Luna:2015paa}. Ultimately, this is possible because the solutions in question are algebraically special, which guides us through the seemingly intractable problem of relating coordinates in a curved spacetime to coordinates in the flat spacetime where the gauge theory solution lives. In particular, the Kerr solution has algebraic type D in the Petrov classification, and moreover it is of Kerr-Schild type, which determines a privileged class of coordinates that can also be thought of as those of a flat spacetime; see \cite{lumonioc} for other exact vacuum type D solutions and \cite{typeNDC} for vacuum type N solutions. The need to worry about gauge choices is absent when dealing only with scattering amplitudes, which are gauge invariant, but an exact relation between explicit classical solutions requires thinking about coordinates. Notice that in order to relate a double copy of classical solutions to the original double copy of scattering amplitudes, one needs to think about the classical solutions perturbatively. This undoes the algebraic magic involved in the exact solutions, but is important in order to show that we are dealing with the same notion of double copy. The equivalence of the notions of double copy has now been established in various ways \cite{Luna:2016due,Goldberger:2016iau,Luna:2016hge,Arkani-Hamed:2019ymq,Huang:2019cja,Kim:2019jwm,Luna:2020adi,Cristofoli:2020hnk,Monteiro:2020plf,Guevara:2020xjx}, with \cite{Monteiro:2020plf} providing a particularly transparent proof directly from scattering amplitudes, using the KMOC formalism \cite{Kosower:2018adc}.

The formulation of the double copy between exact classical solutions to be explored in this paper is the {Weyl double copy} \cite{lumonioc}. It interprets the Weyl curvature of a gravity solution as a double copy of a gauge theory field strength; a scalar field also plays an important role: it is the bi-adjoint scalar field of the scattering amplitudes story. At linearised level, a twistorial interpretation of the Weyl double copy has been proposed \cite{White:2020sfn,Chacon:2021wbr}. A potentially puzzling feature of the double copy of exact solutions is that it deals with Abelian solutions on the gauge theory side; for instance, the single copy of the Schwarzschild solution is the Coulomb solution. This shows that the non-linearity of the gravity solutions in question has the special feature that it can be ``swept under the carpet" by a coordinate transformation, due to their multi-Kerr-Schild property. That this property (particularly double-Kerr-Schild) applies to any type D vacuum solution was shown long ago \cite{Plebanski:1976gy}. In its various guises, Weyl or otherwise, the classical double copy has been a very active research topic; see e.g.~\cite{Ridgway:2015fdl,White:2016jzc,Goldberger:2017frp,Adamo:2017nia,DeSmet:2017rve,Goldberger:2017vcg,Carrillo-Gonzalez:2017iyj,Bahjat-Abbas:2017htu,Luna:2017dtq,Shen:2018ebu, Ilderton:2018lsf,Berman:2018hwd,Bahjat-Abbas:2018vgo,Plefka:2018dpa,Gurses:2018ckx,Cardoso:2016ngt,Cardoso:2016amd,LopesCardoso:2018xes,Andrzejewski:2019hub,Plefka:2019hmz,CarrilloGonzalez:2019gof,Sabharwal:2019ngs,Bah:2019sda,Borsten:2019prq,Borsten:2021hua,Borsten:2021zir,Bahjat-Abbas:2020cyb,Elor:2020nqe,Adamo:2020qru,Alfonsi:2020lub,Keeler:2020rcv,Easson:2020esh,Pasarin:2020qoa,Gumus:2020hbb,Alawadhi:2021uie,Lee:2018gxc,Berman:2020xvs,Lescano:2021ooe,Angus:2021zhy,Cheung:2021zvb,Chacon:2021hfe,Cristofoli:2021vyo,Alkac:2021seh,Farnsworth:2021wvs,Ferrero:2020vww,Alkac:2021bav,Gonzo:2021drq,Saotome:2012vy,Neill:2013wsa}. 

Another line of work that motivates our paper focuses on the asymptotic symmetries of gravity and gauge theory solutions, and has seen a surge in interest in recent years, particularly since the developments reviewed in \cite{Strominger:2017zoo}. Supertranslations have been related via the double copy to electric-magnetic duality \cite{gogopoptn2,Huang:2019cja,Alawadhi:2019urr,Banerjee:2019saj}. Large diffeomorphisms in self-dual gravity and large gauge transformations in self-dual gauge theory have also been related \cite{Campiglia:2021srh}, based on the fact that the self-dual theories provide a simple setting for the double copy \cite{Monteiro:2011pc}. The notions of celestial operators and amplitudes that arose recently are also revealing their own versions of the double copy \cite{Casali:2020vuy,Casali:2020uvr,Pasterski:2020pdk}. Closer to the approach to be taken here, the double copy has been seen to arise in the characteristic value formulation of general relativity, particularly in the example of the Taub-NUT spacetime \cite{gogopoptn2}.  In \cite{gogopoptn2}, it was shown that the Dirac monopole solution can be viewed as a seed solution for the Taub-NUT spacetime in a characteristic value formulation and, moreover, that the proper and large gauge transformations of the monopole solution map onto proper and asymptotic diffeomorphisms of the Taub-NUT solution.

In this paper, we will explore aspects of the classical double copy as seen asymptotically, near null infinity, with a view to gaining new insights into the connections between these various subjects. Bondi coordinates are then a natural choice, as we will explain. We will see how the Weyl double copy relates in a simple manner various quantities that are familiar from the literature of asymptotic symmetries and of the characteristic value formulation. While the Weyl double copy has only been successfully applied to certain algebraically special spacetimes, the fact that a much wider class of spacetimes is {\it asymptotically special} allows us to extend its application, albeit in a restricted asymptotic framework, which we hope will be a stepping stone for a fuller understanding. 

We will consider two examples of spacetimes in detail. One is that of rotating STU supergravity black holes \cite{chcvlupo}, which are algebraically general but asymptotically of type D; the Kerr-Newman solution with equal dyonic charges is a particular case. While a double copy interpretation of STU supergravity is not known, the asymptotic properties of these solutions are sufficient for our purpose.

The other example that we will consider in detail is the C-metric. This was discussed already as an example of the type D Weyl double copy \cite{lumonioc}: the uniformly accelerated black holes are associated via the double copy to the Li\'enard-Wiechert field for uniformly accelerated point charges. The Kerr-Schild double copy is insufficient to deal on its own with this example (even in the multi-Kerr-Schild framework) because of the time dependence, but the Weyl double copy provides a complete prescription. Here, we will revisit the C-metric example of the double copy as seen asymptotically, based on Bondi coordinates, whose construction for the C-metric is not straightforward. A major motivation is the proposed interpretation of the C-metric as a non-linear solution associated to a superrotation \cite{Strominger2016}. This suggests that its single copy can be interpreted analogously as a large gauge transformation, and we will find that this is the case. 

This paper is organised as follows. In section \ref{sec:BondiDC}, we present the `asymptotic' or characteristic value formulation of the Weyl double copy. In section~\ref{sec:AsymWDC}, we apply this to the example of STU black holes, which are only algebraically special asymptotically, and discuss also the simplifications arising for axisymmetric solutions. Section~\ref{sec:LGauge} shows that the superrotation interpretation of the C-metric has a natural single copy analogue, in terms of a large gauge transformation. We conclude with some final comments. There are several appendices containing technical details.

\vspace{.3cm}
{\bf Note added:} While we were finalising the paper, we became aware of parallel work \cite{AdamoKol}, whose goals overlap with ours.

\section{The Weyl double copy in the characteristic value formulation} \label{sec:BondiDC}

As described in appendix \ref{app:spinor}, 
the homomorphism between the Lorentz group and SL$(2,\mathbb{C})$ 
can be used 
to write spacetime tensors as spinors \cite{PenRind}.  The 
Weyl tensor may equivalently be written as a totally symmetric 2-component
spinor object $\Psi_{ABCD}$.  Similarly, a Maxwell 2-form $\cF_{\mu \nu}$ 
may be written as a totally symmetric 2-spinor $\Phi_{AB}$.    
The Weyl double copy is the observation that for type D and N vacuum solutions \cite{lumonioc,typeNDC},\footnote{Some type III examples, in the linearised approximation, were discussed in \cite{White:2020sfn,Chacon:2021wbr}.} the Weyl spinor takes the form
\be
\Psi_{ABCD} =\fft{3c}{S}\, \Phi_{(AB}\, \Phi_{CD)}\label{weylmax},
\ee
where in the type D case
\begin{equation} \label{S}
 \fft{S}{3} = (-2 \Phi^{AB} \Phi_{AB})^{1/4}\,,
\end{equation}
with $c$ a constant designed to absorb parameters. For type N spacetimes, 
$S$ solves the wave equation on the curved background, and also the 
wave equation on a Minkowski background in the case of
non-twisting solutions \cite{typeNDC}.  For type D solutions, in general $S$ only satisfies the wave equation in a flat spacetime, rather than 
the curved background. Indeed, in the type D case, the two equations above imply that
\begin{equation}
 \Box S = -2 (\Psi_2)^{4/3},
 \label{eq:boxS}
\end{equation}
where $\Psi_2$ is a component of the Weyl spinor to be defined momentarily.
Therefore, in order to satisfy the wave equation we need to turn off 
the parameters of the background type D solution, such as the mass, that are 
associated with the curvature, so that $\Psi_2$ then becomes zero,
implying a (at least locally) Minkowski background.

Translating back to tensor language, the 2-component spinor relation (\ref{weylmax}) between the
Weyl curvature and the Maxwell field strength translates into the
tensor relation
\bea
C_{\mu\nu\rho\sigma} + i\, {^*\!C}_{\mu\nu\rho\sigma} &=&
 \fft{c}{S}\, \Big\{
  \cF_{\mu\rho}\, \cF_{\nu\sigma}- \cF_{\mu\sigma}\, \cF_{\nu\rho} +
   2 \cF_{\mu\nu}\, \cF_{\rho\sigma}
+\ft12 (g_{\mu\rho}\,g_{\nu\sigma} - g_{\mu\sigma}\, g_{\nu\rho})\,
    \cF^2\nn\\
&&\quad \ \
- \ft32[(\cF^2)_{\mu\rho}\, g_{\nu\sigma} -(\cF^2)_{\mu\sigma}\, g_{\nu\rho} -
       (\cF^2)_{\nu\rho}\, g_{\mu\sigma} + (\cF^2)_{\nu\sigma}\, g_{\mu\rho}]
\Big\} \label{CplussC}\\
&& \hspace{-13mm} +\fft{i c}{S}\, \Big\{ \ft38(\cF_{\mu\nu}\, {^*\!\cF}_{\rho\sigma} +
  {^*\!\cF}_{\mu\nu}\, \cF_{\rho\sigma} ) -\ft1{16} (g_{\mu\rho}\, g_{\nu\sigma} -
g_{\mu\sigma}\, g_{\nu\rho})\, {^*\!\cF}^{\alpha\beta}\, \cF_{\alpha\beta} -
   \ft1{16} \epsilon_{\mu\nu\rho\sigma}\, \cF^2\Big\}\,,\nn
\eea
where 
\begin{equation}
 \qquad {^*\!C}_{\mu\nu\rho\sigma}= \ft12 \epsilon_{\mu\nu\alpha\beta}\, C^{\alpha\beta}{}_{\rho\sigma}
\end{equation}
and
\begin{equation}
 (\cF^2)_{\mu\nu}=\cF_\mu{}^{\rho}\, \cF_{\nu\rho}, \quad \cF^2=\cF^{\mu\nu}\, \cF_{\mu\nu},\quad {^*\!\cF}^{\alpha\beta}=\ft12\epsilon^{\alpha\beta\rho\sigma}\, \cF_{\rho\sigma}.
\end{equation}
Note that, in general, $c/S$ is complex.  However, if it is real, then 
by taking the real part of (\ref{CplussC})
one has 
\bea
C_{\mu\nu\rho\sigma} &=& \fft{c}{S}\, \Big\{
  \cF_{\mu\rho}\, \cF_{\nu\sigma}- \cF_{\mu\sigma}\, \cF_{\nu\rho} +
   2 \cF_{\mu\nu}\, \cF_{\rho\sigma} 
+\ft12 (g_{\mu\rho}\,g_{\nu\sigma} - g_{\mu\sigma}\, g_{\nu\rho})\,
    \cF^2\nn\\
&&\quad \ \
- \ft32[(\cF^2)_{\mu\rho}\, g_{\nu\sigma} -(\cF^2)_{\mu\sigma}\, g_{\nu\rho} -
       (\cF^2)_{\nu\rho}\, g_{\mu\sigma} + (\cF^2)_{\nu\sigma}\, g_{\mu\rho}]
\Big\}\,.\label{weylmax2}
\eea

We may introduce a Newman-Penrose null frame \cite{NP61} $(\ell, n, m, \bar{m})$, where $\ell$ and $n$ are null vectors such that $\ell \cdot n = -1$ and $m$ is a complex null vector, orthogonal to $\ell$ and $n$, that parametrises the remaining two spacelike directions, so that $m \cdot \bar{m} = 1$.  Therefore, \footnote{We define $v^{(\mu}w^{\nu)}=\frac{1}{2}\left(v^{\mu}w^{\nu}+w^{\mu}v^{\nu}\right)$ and $v^{[\mu}w^{\nu]}=\frac{1}{2}\left(v^{\mu}w^{\nu}-w^{\mu}v^{\nu}\right)$.}
\begin{equation}
 g^{\mu \nu} = -2 \ell^{(\mu} n^{\nu)} + 2 m^{(\mu} \bar{m}^{\nu)},
\end{equation}
or in more familiar language
\begin{equation}
 g_{\mu \nu} = E_\mu^a\, E_\nu^b\, \eta_{ab}, \qquad \eta_{ab} = \begin{pmatrix}
                                                            0 & -1 & \ 0 & \ 0 \\
                                                            -1 & 0 & \ 0 & \ 0 \\
                                                            0 & 0 & \ 0 & \ 1 \\
                                                            0 & 0 & \ 1 & \ 0
                                                            \end{pmatrix}
\end{equation}
with 
\begin{equation}
 E^0 = - n^{\flat}, \quad E^1 = - \ell^{\flat}, \quad E^{m} = \bar{m}^{\flat}, \quad E^{\bar{m}} = m^{\flat}.
\end{equation}
Equivalently,
\begin{equation}
 E_0 = \ell, \quad E_1 = n, \quad E_m = m, \quad E_{\bar{m}} = \bar{m}\,,
\end{equation}
where $E^a=E^a_\mu\, dx^\mu$ and $n^{\flat}= n_\mu\,dx^\mu$, etc., and 
$E_a= E_a^\mu\, \del_\mu$ and $n=n^\mu\, \del_\mu$, etc.
For convenience, we can use the above notation to project any tensor into this null frame.  For example, for any 1-form $X$,
\begin{equation}
 X_0 \equiv \ell^\mu X_{\mu} = - X^1, \quad X_1 \equiv n^\mu X_{\mu} = - X^0, \quad X_m \equiv m^\mu X_{\mu} = X^{\bar{m}}, \quad X_{\bar{m}} \equiv \bar{m}^\mu X_{\mu} = X^{m}.
\end{equation}

A corresponding spinor basis $(\omicron,\iota)$ may be constructed, satisfying
\begin{equation}
 \epsilon_{AB} \,\omicron^A \iota^B = 1,
\end{equation}
where $\epsilon_{AB}$ is a volume form in spinor space, so that
\begin{equation}
 \ell \sim o^A \bar{\omicron}^{\dot{A}}, \qquad n \sim \iota^A \bar{\iota}^{\dot{A}}, \qquad m \sim \omicron^A \bar{\iota}^{\dot{A}}.
\end{equation}
(More precisely, we have $\ell^\mu= E^\mu_a\, \sigma^a{}_{A \dot A}\,
o^A\bar{o}^{\dot A}$, etc., where $\sigma^a{}_{A\dot A}$ is defined in
eqn (\ref{sigdef}).)

In such a null frame, Maxwell and Weyl scalars may be defined by projecting
 the Maxwell field strength and the Weyl tensor into the null frame.  In particular, we define the Maxwell scalars as
\begin{gather}
\Phi_0= \cF_{0m}\, =  \Phi_{AB}\, \omicron^A\, \omicron^B\,,\quad
\Phi_1 =\ft12 (\cF_{01} - \cF_{m \bar{m}})\, =\Phi_{AB}\, \omicron^A\, \iota^B\,, \notag \\[3mm]
\Phi_2= \cF_{\bar m 1}\, =\Phi_{AB}\, \iota^A\, \iota^B\, \label{MaxwellScalars}
\end{gather}
and the Weyl scalars as
\begin{gather}
 \Psi_0=C_{0m0m}\, =\Psi_{ABCD}\, \omicron^A\, \omicron^B\, \omicron^C\, \omicron^D\,,\quad
\Psi_1=C_{010m}\, =\Psi_{ABCD}\, \omicron^A\, \omicron^B\, \omicron^C\, \iota^D\,,\quad \nn \\[3mm]
\Psi_2=C_{0m\bar{m}1}\, =\Psi_{ABCD}\, \omicron^A\, \omicron^B\, \iota^C\, \iota^D\,,\\[3mm] \nn
\Psi_3=C_{101\bar{m}}\, =\Psi_{ABCD}\, \omicron^A\, \iota^B\, \iota^C\, \iota^D\,,\quad
\Psi_4=C_{1\bar{m}1\bar{m}}\, =\Psi_{ABCD}\, \iota^A\, \iota^B\, \iota^C\, \iota^D\,.
\end{gather}
Therefore, translating the Weyl double copy equation \eqref{weylmax} into the null frame constructed above gives
\begin{gather}
\Psi_0 = 3c\, \fft{(\Phi_0)^2}{S}\,,\qquad
\Psi_1 = 3c\, \fft{\Phi_0\,\Phi_1}{S} \,,\qquad
\Psi_2 = c\, \fft{\Phi_0\, \Phi_2 + 2 (\Phi_1)^2}{S} \,,\nn\\[3mm]
\Psi_3 =  3c\,\fft{\Phi_1\, \Phi_2}{S} \,,\qquad
\Psi_4 = 3c\, \fft{(\Phi_2)^2}{S} \,.\label{PsiPhisqrel}
\end{gather}

Having re-expressed the Weyl double copy equation in a null frame, we choose coordinates that will provide a direct relation to a characteristic value formulation of the Einstein equation. We begin by assuming that the spacetime is locally asymptotically flat.~\footnote{By locally asymptotically flat we mean spacetimes that can be put into a Bondi form as described below, but with metric components that are not necessarily regular on the 2-sphere.  
Examples where the components are regular on the sphere
include the Kerr metric and the charged  STU supergravity metrics
that we discuss in this paper.  Examples where there are singularities
on the sphere include the Taub-NUT solution and the C-metric.} Locally asymptotically flat spacetimes provide a mathematical model of an isolated gravitational system that may be 
emitting radiation that is measured by an observer at infinity.  We choose Bondi coordinates $(u,r,x^I=\{\theta, \phi\})$, where $u$ is a timelike coordinate, $r$ is a radial null coordinate and $x^I$ correspond to angular coordinates.  In such a coordinate system, the metric takes the Bondi form~\footnote{In fact, this form of the metric is due to Sachs \cite{sachs}.  The form of the metric that appears in \cite{bondi} is restricted to axisymmetric solutions.  Moreover, the choice of coordinates that was
made in \cite{bondi} is not well-adapted to solution with angular momentum.  
Therefore, except in section \ref{sec:axi}, we shall use the Sachs form even when dealing with axisymmetric solutions.}
\begin{equation} \label{met}
 d  s^2 = - F e^{2 \beta} d  u^2 - 2 e^{2\beta} d  u d  r + r^2 h_{IJ} (d  x^I - C^I d  u)(d  x^J - C^J d  u),
\end{equation}
where we assume the following large-$r$ fall-off conditions for the metric components:
\begin{align} \label{metcomp}
 F(u,r,x^I) &= 1 + \sum_{i=0}^{\infty} \frac{F_{i}(u,x^I)}{r^{i+1}}, \qquad
 \beta(u,r,x^I) = \sum_{i=0}^{\infty} \frac{\beta_{i}(u,x^I)}{r^{i+2}}, \\[3mm]
 C^I(u,r,x^I) &= \sum_{i=0}^{\infty} \frac{C^I_{i}(u,x^I)}{r^{i+2}}, \qquad
 h_{IJ}(u,r,x^I) = \omega_{IJ} + \frac{C_{IJ}}{r} + \frac{C^2\, \omega_{IJ}}{4\, r^2} + \sum_{i=1}^{\infty} \frac{D^{(i)}_{IJ}(u,x^I)}{r^{i+2}} \nn
\end{align}
with $\omega_{IJ}$ the metric on the round 2-sphere.  Note that $C^2=C_{IJ} C^{IJ}$, where we always lower/raise indices on tensors defined on the 2-sphere using $\omega_{IJ}$ and its inverse.  Furthermore, we fix a residual coordinate freedom in the definition of the radial coordinate $r$ by requiring that
\begin{equation} \label{det}
 \textup{det}(h_{IJ}) = \textup{det}(\omega_{IJ}) = \sin^2 \theta.
\end{equation}
Following Ref.\ \cite{sachs}, we may choose a parametrisation of $h_{IJ}$ that 
is adapted to this gauge choice:
\begin{equation}
 2\, h_{IJ} d  x^I d  x^J = (e^{2f} + e^{2g}) d \theta^2 + 4 \sin \theta \sinh (f-g) d \theta d \phi + \sin^2 \theta (e^{-2f}+e^{-2g}) d  \phi^2,
\end{equation}
where 
\begin{equation}
 f(u,r,x^I) = \frac{f_0(u,x^I)}{r} + \sum_{i=2}^{\infty} \frac{f_i(u,x^I)}{r^{i+1}}, \qquad g(u,r,x^I) = \frac{g_0(u,x^I)}{r} + \sum_{i=2}^{\infty} \frac{g_i(u,x^I)}{r^{i+1}}.
\end{equation}
The tensor $C_{IJ}$ is parametrised by $f_0$ and $g_0$, while the
higher $f_i$, $g_i$ (with $i \geq 2$) parametrise the
$D^{(i-2)}_{IJ}(u,x^I)$ tensors.

Assuming appropriate fall-off conditions for the energy-momentum tensor, 
there are equations relating the various metric tensor components; 
see Ref.\ \cite{gogopofake}.  However, here we will keep the discussion general by not assuming any fall-off conditions on the energy-momentum tensor.

Above, we have assumed an analytic expansion in the metric components.  This is a consistent assumption from an initial value problem perspective, in the sense that assuming an analytic fall-off for initial data will guarantee that the evolved solution will remain analytic \cite{sachs}.  However, it does preclude some physically interesting cases \cite{damour, christ}.  Another more general class of consistent fall-offs that one may consider are polyhomogenous spacetimes \cite{chrusciel, Chrusciel:1998he, Kroon:1998tu, ValienteKroon:1999cj}.  Nevertheless, the analytic expansion we assume here will be sufficient for our purposes.

We choose the following null frame associated with the metric \eqref{met}: \cite{gogopofake}
\begin{equation} \label{NPframe}
 \ell = \frac{\partial}{\partial r}, \quad n = e^{-2\beta} \left[\frac{\partial}{\partial u} - \frac12 F \frac{\partial}{\partial r} + C^I \frac{\partial}{\partial x^I} \right], \quad 
 m = \frac{\hat{m}^I}{r} \frac{\partial}{\partial x^I}, 
\end{equation}
where
\begin{equation}
 2\, \hat{m}^{(I} \bar{\hat{m}}^{J)} = h^{IJ}
\end{equation}
with $h^{IJ}$ the matrix inverse of $h_{IJ}.$  In particular, here, we choose 
\begin{equation} \label{Zweibein}
 \hat{m} = \frac{(e^{-f}+i \, e^{-g})}{2}\, \partial_\theta - \frac{i(e^f+i\, e^g)}{2 \sin\theta}\, \partial_\phi.
\end{equation}

In this formulation, the Einstein equation divides into three sets of equations (see, for example, Ref.\ \cite{gogopoptn2}): hypersurface equations, which hold in each $u=\textup{constant}$ hypersurface, evolution equations, which are first order equations in time derivatives, and finally conservation equations that are satisfied on $r=\textup{constant}$ hypersurfaces.  One major advantage of the characteristic formulation of the Einstein equation is that there are no constraint equations, unlike the situation in the initial value formulation.  $C_{IJ}(u,x^I)$ constitutes free data, while $F_0(u_0,X^I)$, $C_1^I(u_0,x^I)$ and $D_{IJ}^{(i)}(u_0,x^I)$ are unconstrained initial data with associated evolution equations.   All other metric functions can then be solved from these functions and their form at time step $u_0+\Delta u$, derived via their evolution equations; see, for example, Ref.\ \cite{gogopofake}.

In such a frame, the Weyl scalars can be written in a $1/r$ expansion,
where they take the form
\begin{equation}
 \Psi_{i} = \mathcal{O}\left(\frac{1}{r^{5-i}}\right).
\end{equation}
More precisely, given our assumptions of analyticity,
one has the expansions
\be
\Psi_i= \sum_{j\ge 0} \psi_i^j\, \fft1{r^{5+j-i}}\,.\label{Psiexp}
\ee
This notable behaviour of the Weyl scalars is known as the peeling property \cite{NP61, wald}.  The $\Psi_i$ have the form
\begin{align}
 \Psi_0 =& \left[ -3 (1+i) (f_2 + i g_2) - \frac{3}{2} \sigma^0 [(\sigma^0)^2 + |\sigma^0|^2 - (\bar{\sigma}^0)^2] +\frac{1}{2} (\bar{\sigma}^0)^3 \right] \frac{1}{r^5} \nn \\[1mm]
 & - \left[6 (1+i) (f_3 + i g_3) \right] \frac{1}{r^6} + \O(\frac{1}{r^7}) 
  \nn \\[3mm]
 \Psi_1 =& \left[ \frac{3(1+i)}{4} (C_1^\theta - i \sin \theta C_1^\phi) + \frac{3}{4} \eth |\sigma^0|^2 + 3 \sigma^0 \eth \bar{\sigma}^0 \right]\frac{1}{r^4} + \O(\frac{1}{r^5}), \nn \\[3mm]
 \Psi_2 =& \frac{1}{2}\left[ F_0 - 2 \sigma^0 \partial_u \bar{\sigma}^0 + \bar{\eth}^2 \sigma^0 - \eth^2 \bar{\sigma}^2 \right]\frac{1}{r^3} + \Big[ 
 F_1 + \frac{(1+i)}{2}\, \bar{\eth}(C_1^{\theta} - i \sin \theta\, C_1^{\phi}) 
  \notag \\[1mm]
& \quad - \frac{(1-i)}{4}\, \eth(C_1^{\theta} + i \sin \theta\, C_1^{\phi})- \frac{3}{4} \eth(\bar{\sigma}^0 \bar{\eth} \sigma^0) + \frac{9}{4} \sigma^0 \bar{\eth} \eth \bar{\sigma}^0 + \frac{1}{4} \bar{\eth} \bar{\sigma}^0 \eth \sigma^0 \Big] \frac{1}{r^4}
 + \O(\frac{1}{r^5}), \nn \\[3mm]
 \Psi_3 =& \eth \partial_u \bar{\sigma}^0 \frac{1}{r^2} + \O(\frac{1}{r^3}), \nn \\[3mm]
 \Psi_4 =& - \partial_u^2 \bar{\sigma}^0 \frac{1}{r} +  \bar{\eth} \eth \partial_u \bar{\sigma}^0 \frac{1}{r^2} + \O(\frac{1}{r^3}), 
 \label{LeadingWeyl}
\end{align}
where 
\begin{equation}
 \sigma^0 = \frac{(1+i)}{2} (f_0 + i g_0)\,.
\end{equation}
Acting on a scalar of spin $n$, we have
\begin{equation}
 \eth \eta = - \frac{(1+i)}{2} \sin^n \theta \left(\frac{\partial}{\partial \theta} - \frac{i}{\sin \theta} \frac{\partial}{\partial \phi} \right) \left(\frac{\eta}{\sin^n \theta} \right).
\end{equation}

In a similar fashion, we may consider a $1/r$ expansion of the Maxwell potential components, which for physically reasonable matter take the form
\begin{gather}
 \cA_u(u,r,x^I) = \sum_{i=0}^\infty \frac{\cA^{(i)}_u(u,x^I)}{r^{i+1}}, 
\qquad \cA_r(u,r,x^I) = \sum_{i=0}^\infty \frac{\cA^{(i)}_r(u,x^I)}{r^{i+2}}, \nn \\[3mm]
 \quad \cA_I(u,r,x^I) = \sum_{i=0}^\infty \frac{\cA^{(i)}_I(u,x^I)}{r^{i}},
\end{gather}
where, again, we assume an analytic form for the dependence of the gauge 
fields on $1/r$.  The analogue of the Bondi gauge in this case is to use gauge freedom to set $\cA_r$ to zero:
\begin{equation}
 \cA \rightarrow \cA - d\Lambda, \qquad \Lambda = \int_{r}^{\infty} \cA_r(u,r',x^I) dr' + \lambda(x^I).
\end{equation}
In this gauge, we have
\begin{gather}
 \cA_u(u,r,x^I) = \sum_{i=0}^\infty \frac{\cA^{(i)}_u(u,x^I)}{r^{i+1}}, \quad \cA_r(u,r,x^I) = 0,  \quad \cA_I(u,r,x^I) = \sum_{i=0}^\infty \frac{\cA^{(i)}_I(u,x^I)}{r^{i}},\label{Aexpansion}
\end{gather}
with a residual gauge freedom parametrised by $\lambda(x^I)$, which 
corresponds to a so-called 
large gauge transformation.  This large gauge transformation is the single copy analogue of the gravitational BMS generator; a statement that we shall make more precise in section \ref{sec:LGauge}.

The corresponding Maxwell field strengths are of the form
\begin{gather}
 \cF_{ur} = -\partial_r \cA_u = \frac{\cA^{(0)}_u}{r^2} + \O(1/r^3), 
 \qquad 
 \cF_{uI} = \partial_u \cA_I - \partial_I \cA_u = \partial_u \cA^{(0)}_I + \O(1/r),  \nn 
 \\[3mm]
 \cF_{rI} = \partial_r \cA_I = - \frac{\cA^{(1)}_I}{r^2} + \O(1/r^3), \qquad \cF_{IJ}= 2 \partial_{[I} \cA_{J]}  = 2 \partial_{[I} \cA^{(0)}_{J]} + \O(1/r).
 \label{Fcompts}
\end{gather}

The Bianchi identity $d \cF=0$ 
is trivially satisfied, while the Maxwell equation
\begin{equation}
 d \star \cF=0
\end{equation}
is equivalent to 
\begin{equation}
 \partial_\mu (\sqrt{-g}\, g^{\mu \rho} g^{\nu \sigma} \cF_{\rho \sigma}) = 0.
\end{equation}
In Bondi coordinates 
\begin{equation}
 \sqrt{-g} = r^2 e^{2\beta} \sqrt{\omega},
\end{equation}
while the inverse metric takes the form
\begin{equation}
 g^{\mu \nu} = \begin{pmatrix}
                0 & -e^{-2\beta} & 0 \\
                -e^{-2\beta} & e^{-2\beta} F  & - e^{-2\beta} C^{J} \\
                0 & - e^{-2\beta} C^{I} & r^{-2} h^{IJ} 
               \end{pmatrix}.          
\end{equation}
In fact, for type D and for non-twisting type N solutions, the Maxwell field appearing in the Weyl double copy satisfies the Maxwell equation also on Minkowski spacetime \cite{lumonioc, typeNDC}.  

Using equations \eqref{MaxwellScalars}, \eqref{NPframe}, \eqref{Zweibein} and \eqref{Fcompts}, we can derive the appropriate Maxwell scalars 
$\Phi_{0}, \Phi_1$ and $\Phi_2$ in a $1/r$ expansion.  For type D 
solutions, the scalar $S$ is then given by
\begin{equation} \label{LeadingScalar}
 \frac{S}{3} = \sqrt{2} \left(\Phi_1^2 - \Phi_0 \Phi_2 \right)^{1/4} =\O(1/r).
\end{equation}

Comparing the $1/r$ expansions of the Weyl scalars \eqref{LeadingWeyl} with the Maxwell scalars and the scalar given by \eqref{LeadingScalar} via the double copy relations \eqref{PsiPhisqrel} gives an asymptotic formulation of the Weyl double copy.

It is important to stress that in formulating the Weyl double copy in the characteristic value formulation, the single copy must be expressed 
in Bondi coordinates on a flat Minkowski background.  However, the Maxwell scalars must be defined with respect to the curved null frame.

\subsection{C-metric and the Li\'enard-Wiechert solution} \label{sec:exCDC}

In this section, we demonstrate the conclusions of the previous section using the C-metric as an example.  The single copy of the C-metric is the analogous Li\'enard-Wiechert solution \cite{lumonioc}. Following the prescription given above in section \ref{sec:BondiDC}, we need to transform the coordinates for the C-metric to Bondi coordinates and also transform the coordinates for the Li\'enard-Wiechert solution to those corresponding to Minkowski in accelerated coordinates (i.e.\ the C-metric with the mass parameter $m$ set to
zero).  In appendix \ref{app:Cmet}, we derive the Bondi form of the C-metric.  In the original C-metric coordinates in which the metric takes the form \eqref{C-metric in xy old G}, the Li\'enard-Wiechert gauge potential reads
\begin{equation}
 \cA= Q\, y\, dt,
\end{equation}
which can easily be seen to give a solution of the
Maxwell equations.  In particular, this is true if we turn off the mass 
parameter, so that the metric just describes Minkowski spacetime in 
accelerating coordinates.  Thus $\cA$ has the interpretation of being the
Li\'enard-Wiechert potential for an accelerating charge.  After rewriting it in
terms of our new Bondi coordinates (see appendix \ref{app:Cmet}), we will simply have that $\cA$ is
given by 
\be
\cA= Q\, \Big(\fft1{\Omega A} - x - T\Big) \Big(dw - \fft{dy}{F(y)}
\Big)\,,
\ee
where we then implement the various substitutions and expansions detailed
in appendix \ref{app:Cmet}. After doing this we find that
\bea
\cA_u &=& -\fft{Q\, x\, \cos\theta}{r\, \sin^2\theta} + 
                 {\cal O}\Big(\fft1{r^2}\Big)\,, \qquad
\cA_r = \fft{Q\, x\, \Gj(x)\, \cos\theta}{A\, r^2\, \sin\theta} +
      {\cal O}\Big(\fft1{r^3}\Big)\,,\nn\\
\cA_\theta &=& -Q\, x\, \csc\theta +  {\cal O}\Big(\fft1{r}\Big)\,, \hspace{10mm}
\cA_\phi = 0\,.
\eea
After a compensating gauge transformation to restore
the $\cA_r =0$ gauge choice, we have
\bea
\cA_u &=& \fft{Q\, \cos\theta}{r\, \sin^2\theta} \left(1-x + G^{3/2} \Gj \right) + 
                 {\cal O}\Big(\fft1{r^2}\Big)\,, \quad
\cA_r = 0\,,\nn\\
\cA_\theta &=& -Q\, x\, \csc\theta +  {\cal O}\Big(\fft1{r}\Big)\,, \hspace{32mm}
\cA_\phi = 0\,.
\eea
These expressions are valid in the general C-metric, but we actually want them
just in the flat spacetime limit, which can be obtained by setting $m$ to zero in the expressions in appendix \ref{app:smallm}.  Thus $x$ and $\Gj(x)$ are then just given 
by the expansions in (\ref{Gjm}) and (\ref{xm}) with $m$ set to 0, and so
we have
\bea
\cA_u &=& \fft{Q\, \cos\theta}{
             r\, \sin^2\theta}\left(1- \frac{u^3 A^3}{(u^2 A^2 +\sin^2\theta)^{3/2}} \right) +
                 {\cal O}\Big(\fft1{r^2}\Big)\,, \quad
\cA_r = 0\,,\nn\\
\cA_\theta &=& -\fft{Q\, u A}{\sqrt{u^2 A^2 +\sin^2\theta}\, \sin\theta} 
         +  {\cal O}\Big(\fft1{r}\Big)\,, \hspace{24mm}
\cA_\phi = 0\,.
\label{LWA}
\eea

  Defining the null tetrad $(\ell,n,m)$ and the scalar components of the Weyl
tensor as in section \ref{sec:BondiDC}, we find that for the C-metric written in
Bondi coordinates as described in appendix \ref{app:Cmet}, we have
\begin{gather}
\Psi_0 = \fft{i  (1-\cos^2\theta\, G\, \Gj^2)^2\, \sqrt{G}\, G'''}{
16 A^3\, \sin^3\theta\, r^5} + {\cal O}(r^{-6})\,,\nn\\
\Psi_1 =\fft{(1+i) \cos\theta\, (1-\cos^2\theta\, G\, \Gj^2)\,
G^{3/2}\, \Gj\, G'''}{16 A^2\, \sin^3\theta\, r^4} + {\cal O}(r^{-5})\,,\nn\\
\Psi_2= -\fft{(1-3\cos^2\theta\, G\, \Gj^2)\, G^{3/2}\, G'''}{
   24 A\, \sin^3\theta\, r^3} + {\cal O}(r^{-4})\,,\nn\\
\Psi_3=-\fft{(1-i) \cos\theta\, G^{5/2}\, \Gj\, G'''}{8\sin^3\theta\, r^2}
  + {\cal O}(r^{-3})\,, \qquad
\Psi_4 = -\fft{i A\, G^{5/2}\, G'''}{4 \sin^3\theta\, r} 
       + {\cal O}(r^{-2})\,.\label{Psileading}
\end{gather}
 
We now make the small-$m$ expansion described in appendix \ref{app:smallm} to give
the leading-order terms in the expansions of the Weyl tensor in 
(\ref{Psileading}):
\begin{gather}
\psi_0^0 = -\fft{3 i\, m\, \sin^2\theta\, (u^2 A^2+1)^2}{4 A^2\, 
  (u^2 A^2 + \sin^2\theta)^{5/2}} + {\cal O}(m^2)\,,\nn\\
\psi_1^0 = -\fft{3(1+i)\, m\, u\, \sin\theta\, \cos\theta\, (u^2 A^2+1)}{
   4 (u^2 A^2 + \sin^2\theta)^{5/2}} + {\cal O}(m^2)\,,\nn\\
\psi_2^0 =  \fft{m\, [(3 u^2 A^2 + 1) \sin^2\theta -2u^2 A^2]}{
2 (u^2 A^2 + \sin^2\theta)^{5/2}} + {\cal O}(m^2)\,,\nn\\
\psi_3^0 = \fft{3(1-i)\, m\, u\, A^2\,\sin\theta\, \cos\theta}{
    2 (u^2 A^2 + \sin^2\theta)^{5/2}} + {\cal O}(m^2)\,,  \quad
\psi_4^0  = \fft{3 i m\, A^2\, \sin^2\theta}{
   (u^2 A^2 + \sin^2\theta)^{5/2}} + {\cal O}(m^2)\,. \label{Psileadingm}
\end{gather}
(See eqn (\ref{Psiexp}) for the definition of the $\psi_i^j$.) 

Now we calculate also the field strength for the Li\'enard-Wiechert potential, and thence the Newman-Penrose scalars
\be
\Phi_0= \cF_{\mu\nu}\, \ell^\mu\, m^\nu\,,\qquad
\Phi_1 =\ft12 \cF_{\mu\nu}\, (\ell^\mu\, n^\nu + \bar m^\mu\, m^\nu)\,,\qquad
\Phi_2= \cF_{\mu\nu}\, \bar m^\mu\, n^\nu\,.
\ee
We find that at leading order in the $1/r$ expansion, these are given by
\begin{gather}
\Phi_0 = -\fft{(1+i)\, Q\,(1-\cos^2\theta\, G\, \Gj^2)\, \sqrt{G}}{
4 A \sin^2\theta\, r^3}
+{\cal O}(r^{-4})\,,\nn\\
\Phi_1 = -\fft{Q\,\cos\theta\, G^{3/2}\, \Gj}{2\sin^2\theta\, r^2}
    +{\cal O}(r^{-3})\,, \qquad
\Phi_2 = \fft{(1-i)\ A\, Q\, G^{3/2}}{2\sin^2\theta\, r} 
  +{\cal O}(r^{-2})\,.
\end{gather}
As with the Weyl scalars, the expressions can be made more explicit in a small-$m$ expansion in which
\begin{gather}
 G(x)=1-x^2 + \O(m), \qquad \Gj(x) =x\, (1-x^2)^{-1/2} + \O(m), \nn \\
 x= u A\, (u^2 A^2 +\sin^2\theta)^{-1/2} + \O(m).
\end{gather}
Thus we have
\begin{gather}
\Phi_0 = -\fft{(1+i) Q \sin\theta\, (u^2 A^2 + 1)}{4 A r^3\, 
    (u^2 A^2 +\sin^2\theta)^{3/2}} +{\cal O}(r^{-4})\,,  \label{Phileading} \\  \nn
\Phi_1 = -\fft{A Q u \cos\theta}{2 r^2\, (u^2 A^2 +\sin^2\theta)^{3/2}} 
+{\cal O}(r^{-3})\,,\qquad
\Phi_2 = \fft{(1-i) A Q \sin\theta}{2 r\, (u^2 A^2 +\sin^2\theta)^{3/2}} 
+{\cal O}(r^{-2})\,. 
\end{gather}

We can see from the results for the Weyl scalars in (\ref{Psileadingm}) and
the Maxwell scalars in (\ref{Phileading}) that a relation of the form seen
in (\ref{PsiPhisqrel}) holds.  If we define
\be
R_0= \fft{\Phi_0^2}{\Psi_0}\,,\quad
R_1= \fft{\Phi_0\, \Phi_1}{\Psi_1}\,,\quad
R_2=\fft{\Phi_0\, \Phi_2 + 2 \Phi_1^2}{3\Psi_2}\,,\quad
R_3=\fft{\Phi_1\,\Phi_2}{\Psi_3}\,,\quad
R_4= \fft{\Phi_2^2}{\Psi_4}\,,
\ee
then to leading order in $1/r$ these are all the same:
\be \label{ScC}
R_a =\frac{S}{3 c} =- \fft{Q^2}{6 m r (u^2A^2 + \sin^2\theta)^{1/2}} + {\cal O}(r^{-2})\,,
\qquad\hbox{for all } a\,.
\ee

From the Weyl double copy, we know that the scalar potential is \cite{lumonioc}
\begin{equation}
 S= \widetilde Q\, (\hat x+y)
\end{equation}
for some constant $\widetilde Q$.  In Bondi coordinates this has the large-$r$ expansion
\be
S= \fft{\widetilde Q\, \sqrt{G}}{A \sin\theta\, r}
-\fft{\widetilde Q\, (2\sqrt{G}\, \Gj + \cos^2\theta\, G G'\, \Gj^2)}{
  4 A^2 \sin^2\theta\, r^2} + {\cal O}(r^{-3})\,\label{Sexpand}
\ee
and in the Minkowski background it reduces to
\be
S= \fft{\widetilde Q}{A r\, (u^2A^2 + \sin^2\theta)^{1/2}} +
{\cal O}(r^{-2})\,.
\ee
Comparing this expression with \eqref{ScC}, we find that the two expressions agree once we choose
\begin{equation}
 c = -\frac{2 m \widetilde Q}{A Q^2}.
\end{equation}

Let us make a clarifying remark. We used the asymptotic Weyl scalars with coefficients given in (\ref{Psileadingm}), which correspond to the linearised order in $m$. On the other hand, the Weyl double copy interpretation of the C-metric is exact \cite{lumonioc}. The linearisation in $m$ is actually equivalent to an alternative, but exact, procedure. In \cite{lumonioc}, double-Kerr-Schild coordinates were used for the exact double copy, and in these coordinates the Weyl spinor is proportional to $m$. The advantage of multi-Kerr-Schild coordinates for the double copy is that they allow us to map the gravitational curved spacetime to a flat spacetime where the gauge field and the scalar live. Asymptotically, however, we are interested in using the Bondi coordinates. So the alternative procedure would be to start with double-Kerr-Schild coordinates for gravity, and then transform these into `flat spacetime Bondi coordinates', which we are using for the gauge field and the scalar. In this way, the Weyl coefficients will indeed be linear in $m$. We chose to proceed as in (\ref{Psileadingm}) for brevity.

\section{Asymptotic Weyl double copy} \label{sec:AsymWDC}

In this section, we consider whether the Weyl double copy holds asymptotically for algebraically general spacetimes.  In the previous section, we formulated the type D and type N Weyl double copy for locally asymptotically flat spacetimes.  In this formulation, the Weyl double copy relation transforms into a tower of relations in a $1/r$-expansion.  The fact that locally asymptotically flat spacetimes become simpler for large $r$, means that it may be possible to satisfy the Weyl double copy relations for leading order(s) in a $1/r$ 
expansion, even when the background is not type D or N and so a full Weyl double copy relation is not known. This could be a stepping stone for a more generic understanding of the classical double copy.

One good reason to expect that it might be possible is the peeling property of the Weyl tensor.  As is evident from the general expressions in \eqref{LeadingWeyl}, the Weyl scalars of a 
locally asymptotically flat spacetime satisfy the fall-offs
\begin{equation}
 \Psi_i = \O(r^{i-5}).
\end{equation}
This means that for large $r$, a generic solution becomes asymptotically type N.  This will mean that the Weyl double copy for type N spacetimes \cite{typeNDC} will apply asymptotically.  Perhaps more interestingly, for a non-radiative spacetime one would expect the radiative parts of the Weyl tensor, parametrised by $\Psi_3$ and $\Psi_4$ in a Bondi null frame, to fall off at a rate much faster than that suggested by the peeling theorem.  As is clear from \eqref{LeadingWeyl}, the leading order terms in $\Psi_3$ and $\Psi_4$ are given by the Bondi news $\partial_u \sigma^0$, which parameterises the flux at null infinity. Therefore, in the absence of flux at null infinity, $\Psi_3$ and $\Psi_4$ fall off at least at the same rate as $\Psi_2$.  If they were to fall off yet faster, it would 
then mean that the leading Weyl scalar at large $r$ would be $\Psi_2$ and 
the spacetime would be asymptotically type D.  In this case, the Weyl double 
copy for type D solutions \cite{lumonioc} applies asymptotically. Here we 
shall consider 
a simple such example of algebraically general black hole solutions that 
become asymptotically type D and therefore obey an asymptotic Weyl double 
copy relation.

\subsection{Rotating STU supergravity black holes} \label{sec:STU}

It is instructive to examine the Weyl double copy equations that applied in
the case of asymptotically-flat type D metrics, but now in the case of 
asymptotically-flat metrics that are algebraically general.  Specifically,
we consider the rotating STU black holes with pairwise-equal charges.  As can be
seen from the discussion of their Petrov type in appendix \ref{app:STUpet}, these black holes
approach type D metrics at large distance.  Thus, we can expect that the 
type D Weyl double copy relations will hold asymptotically in this more 
general setting.

The rotating STU supergravity black holes are solutions of ${\cal N}=2$ supergravity coupled 
to three vector multiplets, and the bosonic sector  comprises the metric,
four $U(1)$ gauge fields and six scalar fields (three dilatonic
and three axionic).  The most general black hole
solutions carry eight independent charges (four electric and four magnetic),
but for our purposes it suffices to consider the case of the black hole
solutions with two electric and two magnetic charges, and where furthermore
these pairs of charges are set equal leaving one electric and one magnetic charge.  The resulting black holes can be 
viewed as solutions of a consistent truncation of the full 
STU supergravity, to a theory 
whose bosonic sector comprises the metric, 2 gauge fields and
two scalars (one dilaton and one axion); the bosonic Lagrangian can be
written as\footnote{There are various ways one can write the bosonic
Lagrangian for STU supergravity, and the truncation we are considering
here, depending upon whether one or more of the gauge fields is dualised.
The Lagrangian we are considering here in eqn (\ref{d4lag2}) is written
in a duality frame in which the rotating black hole metrics in eqns 
(\ref{2charges2}) are supported by one field strength carrying a magnetic
charge and the other carrying an electric charge.  If $F_1$
is dualised, the metrics are then supported by electric charges for
both field strengths.  Details of the relations between various
duality formulations of STU supergravity can be found in
\cite{Cvetic:2021lss}.}
\bea
{\cal L}_4 &=& R\, {*\oneone} -\ft12 {*d\varphi}\wedge d\varphi -
\ft12 e^{2\varphi}\, {*d\chi}\wedge d\chi - \ft12 e^{-\varphi}\, 
 ({*F_{ 1}}\wedge F_{1} + {*F_{2}}\wedge F_{2}) \nn\\
&& - \ft12 \chi\, (F_{1}\wedge F_{1} + 
                           F_{2}\wedge F_{2})\,,\label{d4lag2}
\eea
where $F_1=dA_1$ and $F_2=dA_2$.

The rotating black hole
solution takes the form \cite{chcvlupo}~\footnote{Note that we have placed
bars on all the coordinates here since these are the original coordinates of
the black hole solutions.  We reserve unbarred coordinates for
Bondi coordinates $(u,r,\theta,\phi)$.}
\bea
ds^2 &=& -\fft{\Delta}{W}\, (d\bar{t} - a\, \sin^2\bar\theta\,
d\bar\phi)^2 + W\,  \Big( \fft{d\bar{r}^2}{\Delta} + d\bar\theta^2\Big) 
   + \fft{\sin^2\bar\theta}{W}\, [a\, d\bar{t} - (r_1 r_2 +
   a^2)d\bar\phi]^2\,,\label{2charges1}\\
e^{\varphi} &=& 1 + \fft{r_1\, (r_1-r_2)}{W}= 
\fft{r_1^2 + a^2\, \cos^2\bar\theta}{r_1\, r_2 
                       + a^2\, \cos^2\bar\theta} \,,\qquad
   \chi = \fft{a\, (r_2-r_1)\, \cos\bar\theta}{r_1^2 + a^2\,
     \cos^2\bar\theta}\,,\label{scalars}\\
A_{ 1} &=& \fft{2\sqrt2\, M\, s_1\, c_1\, [a\, d\bar{t} - (r_1 r_2 + a^2)
    d\bar\phi]\, \cos\bar\theta}{W}\,,\nn\\
A_{ 2} &=& \fft{2\sqrt2 M\, s_2\, c_2\, r_1\, (d\bar{t}-a\, \sin^2\bar\theta\,
  d\bar\phi)}{W}\,,\label{2charges2}
\eea
where
\bea
\Delta&=& \bar{r}^2+a^2-2M \bar{r}\,,\qquad W= r_1\, r_2 + a^2\cos^2\bar\theta\,,\nn\\
r_i&=& \bar{r}+ 2M s_i^2\,,\qquad s_i=\sinh\delta_i\,,\qquad 
c_i=\cosh\delta_i\,,
\eea

Here $M$ is a parameter characterising the mass of the black hole, $a$ is
the rotation parameter, and $\delta_1$ and $\delta_2$ are parameters
characterising the magnetic and electric charges, respectively.
If $\delta_1=\delta_2$
the solution reduces to the dyonic Kerr-Newman black hole, which is Petrov type D, with electric and magnetic charges equal to one another.  In fact, the Kerr-Newman metric in Boyer-Lindquist coordinates is derived from these coordinates by the following transformation
\begin{equation}
 \bar{r} \rightarrow \bar{r}-2Ms^2, \qquad M \rightarrow \frac{M}{1+2s^2}, \qquad Q=P=\frac{ \sqrt{2} M s c }{1+2s^2},
\end{equation}
where $s=s_1=s_2$ and $c=c_1=c_2.$

As we show in appendix \ref{app:STUpet}, if the charge parameters $\delta_1$
and $\delta_2$ are
unequal, the black hole spacetime is algebraically general 
(Petrov type I).

Before proceeding, an important clarification is in order. We do not know whether the Kerr-Newman black hole admits a standard double copy interpretation, and the same applies to generic rotating STU black holes. The reason is that while the Kerr-Newman black hole is a solution of Einstein-Maxwell theory, it is not a solution of its ``stringy" analogue Einstein-dilaton-axion-Maxwell theory (``4D heterotic gravity"). The latter theory is well known to arise from a double copy, both in scattering amplitudes \cite{Chiodaroli:2014xia} and in a Kerr-Schild-type classical double copy \cite{Cho:2019ype}. The solution space of 
Einstein-Maxwell theory is not embedded into the 4D heterotic gravity, 
because the dilaton and the axion are generically sourced by the Maxwell field --- unless the latter is null, meaning that both invariants built from the field strength and its dual vanish. So, apart from a restricted set of solutions that are also solutions to 4D heterotic gravity, we do not know whether generic solutions to Einstein-Maxwell theory should allow for a standard double copy interpretation --- a conclusion which extends to STU gravity. In appendix~\ref{app:wdcKN}, we discuss what goes wrong for the Kerr-Newman metric if we naively apply the Weyl double copy. In any case, the example of STU black holes will be sufficient for our purposes of illustrating the asymptotic Weyl double copy, since the solution is asymptotically of a double copy form.

\subsubsection{Asymptotic Weyl double copy relation}
\label{subsubsec:awdc_stu}

In order to investigate whether an asymptotic Weyl double copy relation (\ref{PsiPhisqrel}) exists for the general rotating STU black holes, we first need to construct Bondi coordinates for these solutions.  This is done in appendix \ref{app:STUBon}.  Next we need to find a candidate single copy.  The gauge fields of the solution are natural candidates.~\footnote{In fact, the gauge fields solve a modified Maxwell equation due to scalar couplings: $\nabla_\mu (e^{-\varphi} F^{\mu \nu})=0$.}  Thus from equation (\ref{2charges2}) we shall have a single copy potential of the form
\bea
A= \fft{k_1\, \bar r (d\bar t - a\sin^2\bar\theta\, d\bar\phi)}{
            \bar r^2+a^2\,\cos^2\bar\theta}
  + \fft{k_2\, (a\, d\bar t - (r^2+a^2)d\bar\phi)\cos\bar\theta}{
  \bar r^2 + a^2 \,\cos^2\bar\theta}\,,
\eea
where $k_1$ and $k_2$ are two arbitrary charge parameters.

Next, we may take the Weyl double copy
relations given in (\ref{PsiPhisqrel}), and use them to provide five expressions for the
scalar $S$.  Thus we may define
\bea
\frac{S_0}{c} &=& \fft{3\Phi_0^2}{\Psi_0}\,,\qquad \frac{S_1}{c} = \fft{3\Phi_0\, \Phi_1}{\Psi_1}\,,\qquad
\frac{S_2}{c} = \fft{\Phi_0\,\Phi_2 + 2 \Phi_1^2}{\Psi_2}\,,\nn\\
\frac{S_3}{c} &=& \fft{3\Phi_1\, \Phi_2}{\Psi_3}\,,\qquad 
\frac{S_4}{c}= \fft{3\Phi_2^2}{\Psi_4}\,.\label{S04}
\eea
Following the Weyl double copy prescription, the Maxwell field from which the 
Newman-Penrose scalars $\Phi_0$, $\Phi_1$ and $\Phi_2$ are calculated is 
taken to
be that of an electromagnetic field in the flat-space limit of the
black hole metrics, corresponding to setting the mass and charges to zero.  This
is achieved by taking the quantity $M=0$ in the parameterisation we are
using.  

  After implementing the transformation to the Bondi coordinates, and
introducing the associated Bondi null tetrad frame in the standard way, 
we can calculate the Weyl scalars $(\Psi_0,\Psi_1,\Psi_2,\Psi_3,\Psi_4)$ and
the Maxwell scalars $(\Phi_0,\Phi_1,\Phi_2)$ and substitute into equations (\ref{S04})
in order to find expressions for $S_0$, $S_1$, $S_2$, $S_3$, $S_4$.  The expressions
required need to be calculated to a very high order in the Bondi coordinate expansions, which we
performed in Mathematica and which will not be presented in full detail here.  

   First, we record that the calculation of the invariant quantity 
$I^3-27 J^2$, where $I$ and $J$ are defined in eqn (\ref{IJdef}), leads to
the result that
\bea
I^3-27 J^2 = \fft{9 a^4\, M^{10}\, (s_1^2-s_2^2)^8\, (1+s_1^2+s_2^2)^2\,
\sin^4\theta}{4 r^{26}} +{\cal O}(r^{-27})\,.
\eea
This fall-off is consistent with the results of appendix \ref{app:STUpet}; 
in particular eqns \eqref{STUWeylfalloffs} and \eqref{STUinv}. 
Then, we find that at leading order, all five of the scalars $S_0,\ldots, S_4$ defined
in eqns (\ref{S04}) take the form 
\bea \label{Si}
S_i &=&-\fft{3\sqrt{k_1 + i k_2}}{r} + {\cal O}(r^{-2})\,,
\eea
where we have chosen 
\begin{equation}
 c = -\frac{6 M_B}{(k_1 + i k_2)^{3/2}}
\end{equation}
with 
\bea
M_B = M\, (1 + s_1^2 + s_2^2),
\eea
the Bondi mass of the black hole.
Thus, to leading order, the scalars all agree, and define the ``zeroth copy''
that satisfies the massless Klein-Gordon equation in the flat-space limit.  Moreover, at leading order, the expression in \eqref{Si} is equal to the expression derived from equation \eqref{LeadingScalar}.

At higher order the various expressions differ; we find
\bea
S_1-S_0&=& \fft{M_B \sqrt{k_1+i k_2}\, (s_1^2-s_2^2)^2}{(1+s_1^2+s_2^2)^2\, r^2} +
   \fft{\nu_1\, (s_1^2-s_2^2)^2}{r^3}\ + {\cal O}(r^{-4})\,,\nn\\
S_2-S_0&=& \fft{2M_B \sqrt{k_1+i k_2}\, (s_1^2-s_2^2)^2}{(1+s_1^2+s_2^2)^2\, r^2} +
   \fft{\nu_2\, (s_1^2-s_2^2)^2}{r^3}\ + {\cal O}(r^{-4})\,,\nn\\
S_3-S_0&=& \fft{M_B \sqrt{k_1+i k_2}\, (s_1^2-s_2^2)^2}{(1+s_1^2+s_2^2)^2\, r^2}  +
    \fft{\nu_3\, (s_1^2-s_2^2)^2 + \nu_4\, k_1}{r^3} + {\cal O}(r^{-4})\,,\nn\\
S_4-S_0&=& \fft{0}{r^2} + \fft{\nu_5\, (s_1^2-s_2^2)^2 + \nu_6\, k_1}{r^3} +
{\cal O}(r^{-4})  \,,
\eea
Here, the quantities $\nu_1,\ldots, \nu_6$ are expressions whose precise form 
is unimportant in the present discussion.  We can see that whereas in the general
case where the two charges are unequal ($s_1\ne s_2$) the scalars $S_i$ agree only 
at the leading $1/r$ order, the agreement extends to the first sub-leading order
(i.e.\ order $1/r^2$) in the case $s_1=s_2$ (the Kerr-Newman black hole) where the
metric is of Petrov type D. 

If we set $M=0$, the $S_i$ become equal, to the orders calculated, 
to the expression derived from \eqref{S} on the Minkowski background, and they satisfy the wave equation on the Minkowski background.

\subsection{Axisymmetric Weyl double copy} \label{sec:axi}

We have discussed above how to express the Weyl double copy \eqref{PsiPhisqrel} asymptotically starting from the Bondi form of the metric \eqref{met}. The relation becomes lengthy if one attempts to write it down in terms of the $1/r$ expansions for the metric and the gauge field described in that section. For illustrative purposes, particularly regarding the discussion of the next section, we will now consider the restriction to the axisymmetric case, which simplifies the map considerably. The original axisymmetric Bondi metric reads \cite{bondi}
\begin{equation}
\begin{split}
d s^2 =-\left(\frac{V}{r}e^{2\beta}-U^2r^2e^{2\gamma}\right)d u^2-2e^{2\beta}d u d r 
 \hspace{4cm} \\
-2Ur^2e^{2\gamma}d u d \theta+ r^2(e^{2\gamma}d\theta^2+e^{-2\gamma}\sin^2\theta\,d \varphi^2)~,\label{AxiSymBondiMetric}
\end{split}
\end{equation} 
with fall-off conditions
\begin{align}
\gamma(u,r,x^I)&=\frac{c(u,x^I)}{r}+\O(r^{-3})~,\\
\beta(u,r,x^I)&=-\frac{c(u,x^I)^2}{4\,r^2}+\O(r^{-3})~,\\
U(u,r,x^I)&=-(c_{,\theta}+2c \cot\theta)\frac{1}{r^2}+({2\,N(u,x^I)}+3 c\, c_{,\theta}+4 c^2 \cot\theta)\frac{1}{r^3}+\O(r^{-4})~,\\
V(u,r,x^I)&=r-2M_B(u,x^I)+\O(r^{-1})~.
\end{align}
This Bondi form is a subclass of the more general form \eqref{met}, to which it is related by
\begin{align}
F&=\frac{V}{r} \quad\Rightarrow \quad F_0\ = - 2 M_B~,\\
C^I&=(U,0)~, 
\\
h_{IJ}&=\begin{pmatrix}
e^{2\gamma}&0\\
0& e^{-2 \gamma}\sin^2\theta
\end{pmatrix}
\ \Rightarrow \quad C_{\theta\theta}\ =2\,c~.
\end{align}
For the gauge field, the axial symmetry allows us to set $\mathcal{A}_\phi=0$ in \eqref{Aexpansion}. Finally, we introduce an additional simplification, by taking the scalar $S$ to be real in (\ref{CplussC}), which leads to \eqref{weylmax2}; this applies for instance to the C-metric. \footnote{We also assume here the reality of the couplings-absorbing constant $c$ in \eqref{weylmax2}. It should not be confused with the metric function $c$ in the present section, so here we will choose that constant to take the numerical value $1/3$.} 
We can plug these expressions directly into \eqref{weylmax2} and compare the components to obtain the neat relations
\begin{subequations}\label{eq: Bondi weyl double copy}
\begin{align}
\frac{(\mathcal{A}^{(0)}_{\theta,u})^2}{2\,S}&=-c_{,uu}\,~,\label{eq: cuu as AA}\\[1em]
 \frac{\mathcal{A}^{(0)}_{\theta,u}\,\mathcal{A}^{(1)}_u}{2\,S}&= -\frac{\partial_\theta (\sin^2\theta\,c_{,u})}{\sin^2 \theta} \,~,\label{eq: cqu +2cotq cu as AA}\\[1em]
\frac{(\mathcal{A}^{(1)}_u)^2+\mathcal{A}^{(1)}_\theta\,\mathcal{A}^{(0)}_{\theta,u}}{6\, S}&=-M-c\,c_{,u}\,~,\label{eq: M as AA}\\[1em]
\frac{\mathcal{A}^{(1)}_u\,A^{(1)}_\theta}{6\,S}&=N\,~.\label{eq: N as AA}
\end{align}
\end{subequations}
In principle, these expressions could be used to obtain (up to constants of integration) a metric tensor from any axisymmetric gauge potential in Bondi gauge. However, the system \eqref{eq: Bondi weyl double copy} is over-complete, which gives rise to the integrability condition
\begin{equation}
\partial_u \left(\frac{\cA^{(1)}_u\cA^{(0)}_{\theta,u}}{S}\right)= \frac{1}{\sin^2 \theta} \partial_\theta \left( \sin^2 \theta \frac{(\cA^{(0)}_{\theta,u})^2}{S}\right)~.\label{eq:IntegraCondition}
\end{equation}
Alternatively, this equation can be obtained by imposing the vacuum Bianchi identities, $\nabla^{\mu}C_{\mu\nu\rho\sigma}=0$, on \eqref{weylmax2}, assuming that the gauge field satisfies the Maxwell equations.

The expressions above can easily be checked to hold for the example of the C-metric.


\section{Asymptotic symmetries and the Weyl double copy} \label{sec:LGauge}

The classical double copy is fundamentally about relating solutions in gravity and gauge theory.  An important aspect of both gravitational and gauge theories is their symmetry structure.  In gravity, this is given by diffeomorphisms, while in gauge theory it is gauge transformations.  Proper diffeomorphisms and gauge transformations, while not physical in the sense that they parametrise redundancies in the description of the same physics, are important aspects of the theories.  In addition, there exist also improper transformations, which generally change boundary conditions and hence are physical.  Any claim towards a new understanding of such theories ought to give some insight into how the respective symmetry structures arise and how they relate to one another. No fully general relation has been found yet, but much progress has been made on both global and linearised local symmetries, e.g.~\cite{Borsten:2013bp,Anastasiou:2014qba,Anastasiou:2015vba,Anastasiou:2016csv,Anastasiou:2018rdx}, and more recently on asymptotic symmetries  \cite{gogopoptn2,Huang:2019cja,Alawadhi:2019urr,Banerjee:2019saj,Pasterski:2020pdk,Campiglia:2021srh}.

The difficulty arises from the nature of the double copy formulations.  The Kerr-Schild double copy \cite{Monteiro:2014cda} relies on a particular coordinate system (gauge-fixed frame) in the gravitational theory and a gauge-fixed Yang-Mills field for the single copy, leaving no room for symmetry transformations.  In contrast, the Weyl double copy map \cite{lumonioc} still relies on having a map between coordinates in gravity and those in gauge theory, but the Maxwell field strength is gauge invariant in nature.
The leading-order asymptotic Weyl double copy is completely insensitive to symmetry transformations.  This can be easily checked for the simplified case of axial symmetry and real scalar considered in \eqref{eq: Bondi weyl double copy}.

Despite these challenges, some progress has been made on improper or asymptotic transformations.  In \cite{gogopoptn2}, it 
was shown that (proper and improper) supertranslations on the Taub-NUT background correspond to ({proper and improper/large) gauge transformations of the Dirac monopole field.  This relation, which has been explored also in \cite{Huang:2019cja,Alawadhi:2019urr}, relied heavily on the time-independence of the background.  Recently, a more general relation has been given in the context of the self-dual sectors of the respective theories in a light-cone formulation \cite{Campiglia:2021srh}.  Within the self-dual sector, all fields within both the gravitational and Yang-Mills theory can be described in terms of scalars, and this formulation helps in relating the symmetries; in fact, it is known to help make the double copy fully manifest at the level of the equations of motion or the Lagrangian \cite{Monteiro:2011pc}.
It is not clear how similar ideas can be used more generally.

In this section, we study this problem by focusing on the particular example of the C-metric and its single copy, the associated Li\'enard-Wiechert solution.   In \cite{Strominger2016}, it was argued that there is a superrotation embedded in the C-metric. From a double copy perspective, this result indicates that there ought to be an analogous large gauge transformation embedded in the Li\'enard-Wiechert potential. Indeed, we find that the superrotation on the gravitational side maps to a large residual gauge transformation at leading order.

\subsection{The C-metric as a superrotation}

First, we review the results of \cite{Strominger2016}, where it is argued that superrotations are to be viewed as a ``memory effect" related to the appearance/disappearance of cosmic strings piercing null infinity. The C-metric, as an exact solution, provides a non-linear realisation for such a process; see figure  \ref{fig:PenDiagrams}.


\begin{center}
\begin{figure}[ht]\centering
\includegraphics[scale=0.8]{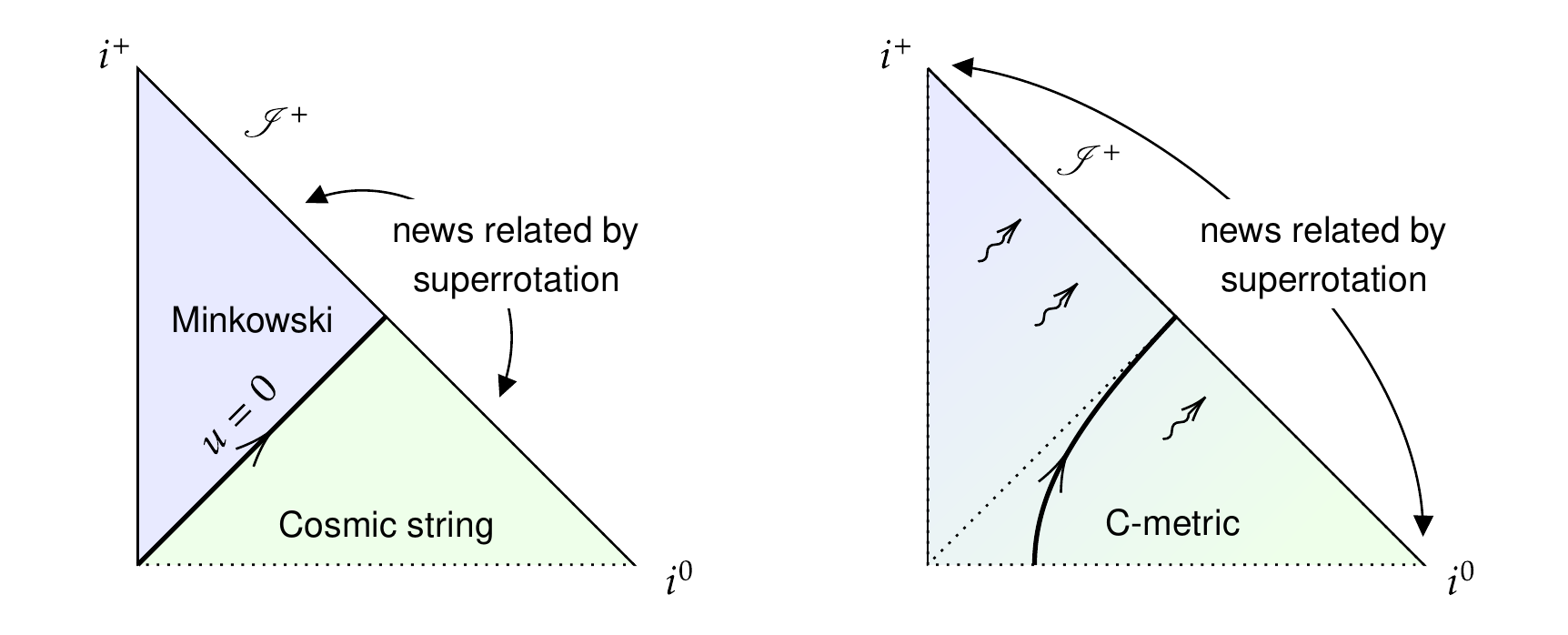}
\caption{A portion of the Penrose diagram for the snapping cosmic string considered in \cite{Strominger2016} is shown on the left. The green area represents the infinite cosmic string metric, which is glued to a Minkowski patch (in blue) along $u=0$. The trajectory of the endpoints is represented by the thick line. On the right is the corresponding picture for the (linearised) C-metric. In this case, we have radiation approaching $\scri^+$. 
Note that the true Penrose diagram for the C-metric is more involved \cite{Griffiths2006}. }\label{fig:PenDiagrams}
\end{figure}
\end{center}

Consider the Bondi news of the C-metric~\footnote{In \cite{Strominger2016}, the solution chosen was the charged C-metric, since it admits a thermodynamic interpretation: a relation between $A$ and $q$ can be imposed such that the event and acceleration horizons have equal surface gravities, hence equal temperatures. This is not needed for our purposes. In fact, we will set $q=0$ since the Weyl double copy is only known to apply in vacuum; see related comments just above section~\ref{subsubsec:awdc_stu}. On the other hand, the single copy gauge field to be considered later, which lives in flat spacetime, coincides precisely with the gauge field of the charged C-metric after taking $m\to0$.}
\begin{equation}
N_{\theta \theta} \equiv \partial_u C_{\theta\theta}=- \frac{1}{4 \sin ^2\theta} \left(4 + 2 \kappa ^2 G(x) G''(x)-\kappa ^2 G'(x)^2\right),\label{eq:news}
\end{equation}
where $\kappa$ parametrises the deficit angle.  Note that in appendix \ref{app:Cmet}, we have normalised $\kappa$ to one.  The expression above can be derived by making the $\kappa$ dependence explicit in the metric \eqref{C-metric in xy old G}, i.e.\ multiplying the $d\phi^2$ term by $\kappa^2$, and using equations \eqref{xdef} and \eqref{Gjdef}, as well as the expression for $C_{\theta \theta}$ in \eqref{CCF}.

We are interested in the limits $u\to\pm\infty$. Recall that $x$ is implicitly defined in \eqref{xdef} in terms of the Bondi $u$ and $\theta$ coordinates, and so using Hong-Teo's parametrisation for $G(x)$, in equation \eqref{chargedHT}, these limits correspond simply to $x=\pm 1.$  The asymptotic structure of \eqref{xdef} gives
\begin{subequations}
\label{eq:limits of u}
\begin{align}
x&\to 1-\frac{\sin^2\theta}{2A^2\kappa^2\,(1+2Am)^3} \ \frac{1}{u^2}+\O(u^{-4}) \quad \text{as} \quad u\to\infty~,
\\
x&\to -1+\frac{\sin^2\theta}{2A^2\kappa^2\,(1-2Am)^3} \ \frac{1}{u^2}+\O(u^{-4}) \quad \text{as} \quad u\to-\infty~.
\end{align}
\end{subequations}

Fixing $\kappa$ so that the segment between the two black holes is regular gives
\begin{equation}
\kappa=\frac{2}{|G'(1)|}=\frac{1}{1+2\,A\,m}~.
\end{equation}
This then implies that
\begin{subequations}\label{eq:newslimits}
\begin{align}
\lim_{u\to-\infty}N_{\theta\theta}&= -\frac{8\, A\,m}{(1+2\,A\,m)^2}\ \frac{1}{\sin^2\theta}~,\label{eq:NewsPast}
\\
\lim_{u\to+\infty}N_{\theta\theta}&=0~.\label{eq:NewsFuture}
\end{align}
\end{subequations}

The $u\to\pm\infty$ limits of the Bondi news, \eqref{eq:NewsFuture} and \eqref{eq:NewsPast}, are related by a superrotation. To show this, we start from a Minkowski background, for which $N_{\theta\theta}=0$. Superrotations are generated by 2-dimensional vector fields $Y^I$ that are conformal Killing vectors of the celestial sphere and independent of $u$ and $r$.\footnote{In other words, they must generate transformations of the type
\begin{equation*}
z\to f(z)~,\qquad z=e^{i\phi}\cot\frac{\theta}{2}~,
\end{equation*}
with $f(z)$ a holomorphic function.}
We are interested in the subgroup that preserves $\partial_\phi$ as a Killing vector of the metric. This restricts the superrotations to a three-parameter subgroup
\begin{equation}
Y^\theta=\left(\beta+\mu\ln\tan\frac{\theta}{2}\right)\,\sin\theta~,\qquad Y^\phi=\mu\,\phi+\vartheta~.
\end{equation}
Setting $\vartheta=0$ and $\beta=0$, the effect on the Bondi news of the 
flat Minkowski metric is 
\begin{equation}
N_{\theta\theta}\to- \mu\,\frac{1}{\sin^2\theta}~.
\end{equation}
A comparison with \eqref{eq:NewsPast} reveals that 
the two limits of the Bondi news of the C-metric  \eqref{eq:newslimits} are related by the superrotation 
\begin{equation}
Y^\theta=\frac{8\, A\,m}{(1+2\,A\,m)^2} \,\sin\theta\;\ln\tan\frac{\theta}{2}~,\qquad
 Y^\phi=\frac{8\, A\,m}{(1+2\,A\,m)^2}\,\phi~.\label{eq:ConformalKillingSuperrot}
\end{equation}

\subsection{The Li\'enard-Wiechert potential as a large gauge transformation}
Having made the case for the C-metric as a superrotation, it is reasonable to expect that its single copy, the Li\'enard-Wiechert potential, has a similar interpretation in terms of a large gauge transformation.

In order to investigate this, we need to put the Li\'enard-Wiechert potential on a background in which Minkowski spacetime is written in Bondi coordinates and in a gauge in which $\cA_{r}=0.$  We have already done this in section \ref{sec:exCDC}, with the appropriate expression given by \eqref{LWA}.  In these coordinates, a large gauge transformation corresponds to 
\begin{equation}
\cA^{(0)}_\theta\to \cA^{(0)}_\theta-\partial_\theta\lambda(\theta)~.\label{eq:resudialgauge}
\end{equation}
These gauge transformations are the electromagnetic 
analogues of the BMS asymptotic symmetries \cite{He2014a}. 

As we did in the previous section, we will compare the two limits of the gauge potential
\begin{equation}
\cA_\pm := \lim_{u\to\pm\infty}\cA~.
\end{equation}
Taking this limit in the expression for the gauge potential \eqref{LWA} gives 
\begin{subequations}\label{eq:limitsofA}
\begin{align}
\cA_+&=
\O(r^{-3})\,d  u
+
\left(
	-\frac{q\,}{\sin\theta}
	+\O(r^{-2})
\right)\,d \theta~,
\\[1em]
\cA_-&=
\O(r^{-3})\,d  u
+
\left(
	\frac{q\,}{\sin\theta}
	+\O(r^{-2})
\right)\,d \theta~,
\end{align}
\end{subequations}
Taking the difference between these gauge potentials we find that
\begin{equation}
(\cA_+-\cA_-)\Big|_{r^0}=-\frac{2q}{\sin\theta}d  \theta=d \lambda(\theta)~,
\end{equation}
where
\begin{equation}
\lambda(\theta)=-2q\,\ln\tan\frac{\theta}{2}~.\label{eq: lambda theta gauge}
\end{equation}
This is indeed a large gauge transformation; compare with \eqref{eq:resudialgauge}.
Note the similarities between  \eqref{eq:ConformalKillingSuperrot} and \eqref{eq: lambda theta gauge}: it is not just that the solutions can be thought of as large diffeormorphisms or gauge transformations, but the corresponding parameters are also related. It would be interesting to find a fully general relation, beyond the example studied here.

\section{Discussion}

We have taken steps towards providing an asymptotic understanding of the classical double copy with the goal of studying the asymptotic symmetries and of making a connection with recent advances in celestial (i.e.~flat-space) holography.

We have shown how the Weyl double copy can be formulated asymptotically, in the neighbourhood of null infinity, where it applies to a wider class of spacetimes, including algebraically general ones. The next natural step would be to understand this formulation at sub-leading orders, i.e.~moving from asymptotic infinity into the bulk. This may assist us in generalising the Weyl double copy beyond algebraically special solutions in an appropriate expansion.

It would also be interesting to understand how our formulation fits in with the story of conformally primary metrics on the celestial sphere and their double copy interpretation \cite{Pasterski:2020pdk}, where the C-metric still provides a puzzling example.

\section*{Acknowledgements} 

We thank the authors of \cite{AdamoKol} for sharing their paper prior to publication. HG would like to thank Queen Mary University of London, and MG and CNP would like to thank the Albert Einstein Institute, Potsdam, for hospitality during the course of this work.  HG is supported by the ERC Advanced Grant Exceptional Quantum Gravity (Grant No.\ 740209).  MG and RM are supported by Royal Society University Research Fellowships. DPV is supported by a Royal Society studentship. CNP is partially supported by DOE grant DE-FG02-13ER42020.

\appendix

\section{Four-dimensional spinor formalism} \label{app:spinor}

  Following the conventions of 
\cite{lumonioc}, define $\sigma^a_{A\dot A}$ and 
$\tilde\sigma^{a\, \dot A A}$ by
\be
\sigma^a= \ft1{\sqrt2}\, (\oneone,\sigma^i)\,,\qquad
\tilde\sigma^a = \ft1{\sqrt2}\, (\oneone,-\sigma^i)\,,\label{sigdef}
\ee
where $\sigma^i$ are the standard Pauli matrices. ($a,b,c,\ldots$ 
are tangent-frame vector indices.)  One also defines
\be
\sigma^{ab}_{AB} = \sigma^{[a}_{A\dot A}\, \tilde\sigma^{b]\,\dot A C}\,
\epsilon_{CB}\,,
\ee
where $\epsilon^{12}=+1$.  Indices are raised and lowered using the epsilon
symbols:
\be
\psi^A=\epsilon^{AB}\, \psi_B\,,\qquad \psi_A=\psi^B\, \epsilon_{BA}\,,
\ee
and similarly for dotted indices.

  A Maxwell field strength $\cF_{ab}$ is written in terms of a symmetric
bi-spinor $\Phi_{AB}$, defined by
\be
\Phi_{AB}= \ft12 \cF_{ab}\, \sigma^{ab}_{AB}\,.
\ee
This has the complex conjugate 
\be
\bar\Phi_{\dot A\dot B}= \ft12 \cF_{ab}\, \bar\sigma^{ab}_{\dot A\dot B}\,,
\ee
where the components of $\bar\sigma^{ab}_{\dot A\dot B}$ are just the complex
conjugates of the components of $\sigma^{ab}_{AB}$.  Inversely, one has
\be
\cF^{ab}= \Phi_{AB}\, \sigma^{ab\, AB} + \bar\Phi_{\dot A\dot B}\,
\bar\sigma^{ab\, \dot A\dot B}\,.
\ee
A straightforward result that follows from the definitions is that
\be
\Phi^{AB}\, \Phi_{AB} = \ft14(\cF^{ab}\, \cF_{ab} + 
                        {i\, ^*\!\cF}^{ab}\, \cF_{ab})\,.
\ee

  In a similar fashion, one defines from the Weyl tensor $C_{abcd}$ the
Weyl spinor
\be
\Psi_{ABCD}= \ft14 C_{abcd}\, \sigma^{ab}_{AB}\, \sigma^{cd}_{CD}\,,
\ee
and its complex conjugate $\bar\Psi_{\dot A\dot B\dot C\dot D}$.  These
quadri-spinors are totally symmetric, by virtue of the symmetries of
the Weyl tensor.  Conversely, one has
\be
C^{abcd}= \sigma^{ab}_{AB}\, \sigma^{cd}_{CD}\, \Psi^{ABCD} +
   \bar\sigma^{ab}_{\dot A\dot B}\, \bar\sigma^{cd}_{\dot C\dot D}\, 
  \bar\Psi^{\dot A\dot B\dot C\dot D}\,.
\ee

   Noting that the symbols $\sigma^{ab}_{AB}$ obey the relation
\be
\sigma^{ab}_{AB}\, \sigma^{cd\, AB} = \ft18 (\eta^{ac}\,\eta^{bd} -
  \eta^{ad}\, \eta^{bc} + i \epsilon^{abcd})\,,
\ee
(and {\it mutatis mutandis} for the complex conjugates 
$\bar\sigma^{ab}_{\dot A\dot B}$), it follows that
\be
\Psi^{ABCD}\, \Psi_{ABCD} = \ft14 (C^{abcd}\, C_{abcd} +
 i\, {^*\!C}^{abcd}\, C_{abcd})\,.
\ee
Note that the Weyl tensor satisfies the property 
\begin{equation}
 {^*\! C}_{abcd}= {C^*}_{abcd}
\end{equation}
and hence also ${{^*\! C^*}_{abcd}=- C_{abcd}}$.

\section{C-metric in Bondi coordinates} \label{app:Cmet}

In this appendix, we shall derive the Bondi form of the C-metric.  There exists a variety of papers addressing the radiative properties of the C-metric in the literature. An early attempt to study the C-metric using the Bondi method was carried out by Bi\v{c}\'ak \cite{Bicak1968}, who gave an expression for the Bondi news (see also \cite{Pravda2000, Sladek2010a, Tomimatsu1998}).  Other studies of the asymptotic properties of the C-metric include \cite{Farhoosh1980, Ashtekar1981, Bini2005, Griffiths2006, Maluf2007}.  Here, we provide a systematic procedure for deriving Bondi coordinates for the C-metric to any desired order in a $1/r$ expansion.

The most common form of the C-metric is
\begin{equation}
d  s^2= \frac{1}{A^2(x+y)^2}\left[-F(y)d  t^2+\frac{d  y^2}{F(y)}+\frac{d  x^2}{G(x)}+G(x)d \phi^2\right]~,
\label{C-metric in xy old G}
\end{equation}
where
\begin{equation}
G(x)=1-x^2-2m\,A\,x^3~,\qquad F(y)=-G(-y)\,.
\end{equation}
Here,
$0<2 \, A\, m < 1$, \ $ y \in (-x,\infty)$ and $x\in (x_2,x_3)$, 
with $x_2$ and $x_3$ the largest two roots of $G(x)$. 

An alternative form of the C-metric, given by Hong and Teo \cite{Hong2003}, 
has the same metric \eqref{C-metric in xy old G} but with a different form  for the functions $F(y)$ and $G(x)$ with a simplified root structure:
\begin{equation}
G(x)=(1-x^2)(1+2\,m\,A\,x)~,\qquad F(y)=-G(-y)~,
\end{equation}
where $ y \in (-x,\infty)$ and $x\in (-1,1)$, $G(x) > 0$.  This patch covers both the static and asymptotic regions, which is given by the limit $x\rightarrow-y$ \cite{Griffiths2006}.

In addition, the charged C-metric solution is given by same metric \eqref{C-metric in xy old G}, but now with
\begin{equation} \label{chargedHT}
G(x)=(1-x^2)(1+r_+\, A\,x)(1+r_-\, A\,x)~,\qquad F(y)=-G(-y)~, 
\end{equation}
where $r_\pm = m \pm \sqrt{m^2 - q^2}$ and $0< r_- A < r_+ A <1.$  The coordinate ranges are the same as those of the uncharged Hong-Teo coordinate system.
 
Given the fact that the form of the metric is the same in both coordinate systems, one can derive a Bondi form from both coordinate systems 
in one go; this is what we now proceed to do.  
We begin by relabelling $x$ as $\hat{x}$ and defining coordinates $\Omega$ and
$w$ by
\begin{equation}
 \Omega=\frac{1}{A(\hat{x}+y)}, \quad w=t+\int d  y/F(y)\,.
\end{equation}
The metric now takes the form
\begin{equation}
d  s^2=\Omega^2 \left[-F(y)d  w^2-\frac{2}{A\Omega^2}d  w d  \Omega -2d  w d  \hat{x}+\frac{d  \hat{x}^2}{G(\hat{x})}+G(\hat{x})d \phi^2\right]\,,
~\label{C-metric in wome}
\end{equation}
with the understanding that $y$ is given in terms of $\hat x$ and $\Omega$
by
\be
y= -\hat x + \fft{1}{A\,\Omega}\,.\label{ydef}
\ee

Now, replacing $\hat x$ by $\alpha$, with
\be
 \fft{d  \hat x}{G(\hat x)} = d  w - \fft{d  \alpha}{\sin\alpha}\label{cons}
\ee
gives
\bea
d  s^2 = -\Omega^2\, [F(y) + G(\hat x)]\, d  w^2 -
  \fft{2 d  w\, d  \Omega}{A} + \fft{\Omega^2\, G(\hat x)}{\sin^2\alpha}\,
(d  \alpha^2 + \sin^2\alpha\, d  \phi^2)\,.\label{cmet1}
\eea
To summarise, we have transformed the C-metric into coordinates given by $(w, \Omega, \alpha, \phi)$ with an auxiliary coordinate $\hat{x}$ given implicitly in terms of the other coordinates via equation \eqref{cons}.

We can now proceed perturbatively, order by order in an inverse distance 
expansion, to put the metric into the Bondi form with coordinates $(u,r,\theta, \phi)$.  However, before we do this, it will prove useful to define a new auxiliary coordinate $x$ implicitly in terms of the new coordinates
$u$ and $\theta$ by
\be
u = \fft{\Gj[x]\, \sin\theta}{A}\,,\label{xdef}
\ee
where
\be
\Gj(x)\equiv \int^x \fft{d  x'}{G(x')^{3/2}}\,.\label{Gjdef}
\ee
The coordinate transformation will define the old coordinates in terms of the new coordinates, i.e.\ we have
\begin{equation}
 w=w(u,r,\theta), \quad \Omega=\Omega(u,r,\theta), \quad \alpha= \alpha(u,r,\theta)
\end{equation}
as well as a relation between the old auxiliary coordinate $\hat{x}$ and the new coordinates
\be
\hat x = x + T(r,x,\theta)\,.\label{xhatx}
\ee
Writing these relations in terms of $1/r$ expansions
\bea
\Omega &=& \bar g_1(x,\theta) \, r + g_0(x,\theta) +
  \fft{g_1(x,\theta)}{r} + \fft{g_2(x,\theta)}{r^2} + \cdots\,,\nn\\
\alpha &=& \theta + \fft{\bar h_1(x,\theta)}{r} + 
\fft{h_0(x,\theta)}{r^2} + \fft{h_1(x,\theta)}{r^3} +\cdots\nn\\
w &=& \bar f_1(x,\theta) + \fft{f_0(x,\theta)}{r} +
  \fft{f_1(x,\theta)}{r^2}+ \fft{f_2(x,\theta)}{r^3} +\cdots\nn\\
T&=& \fft{k_0(x,\theta)}{r}  + \fft{k_2(x,\theta)}{r^2} + 
  \fft{k_3(x,\theta)}{r^3} +\cdots\,,\label{expand}
\eea
we proceed in a systematic manner, requiring that the metric expressed in
terms of the new coordinates $(u,r,\theta,\phi)$ have the Bondi form \eqref{met} and satisfying the fall-offs \eqref{metcomp} as well as the gauge condition \eqref{det}.  Note that in addition, we have the constraint that the old auxiliary coordinate $\hat{x}$ must satisfy equation \eqref{cons}.

We have chosen to define $x$ so that it is equal to $\hat{x}$ up to a perturbative correction $T=\O(1/r)$, as can be seen from 
eqns (\ref{xhatx}) and (\ref{expand}).  
This means that $G(\hat{x})$ and $G(y)$ can simply be written in terms of $x$ using a Taylor expansion.  In particular,
\begin{equation}
 G(\hat{x}) = G(x)+T G'(x)+\frac{1}{2} T^2 G''(x)+ \frac{1}{6} T^3 G^{(3)}(x) + \frac{1}{24} T^4 G^{(4)}(x),
\end{equation}
where primes denote derivatives with respect to $x$, 
and we have used the fact that $G$
is a quartic polynomial in its argument.

  We now proceed order by order in inverse powers of $r$,
by plugging the expansions 
(\ref{expand}) into the metric (\ref{cmet1}) and the constraint (\ref{cons}).
First, we solve equation (\ref{cons})
at order $r^0$, obtaining first-order differential equations for 
$\bar f_1(x,\theta)$ whose solution is
\be
\bar f_1(x,\theta)= Gi(x) + \log\tan\ft12\theta\,,\qquad Gi(x)\equiv 
  \int^x \fft{dx'}{G(x')}\,.
\ee
 From this point on, all the expansion coefficients can be solved purely
algebraically, according to the following scheme:

\begin{itemize}

\item[(a)] Solve equation (\ref{det}) at order $r^0$ for $\bar g_1(x,\theta)$.

\item[(b)] Solve $g_{r\theta}=0$ at order $r^0$ for $\bar h_1(x,\theta)$.

\item[(c)] Solve $g_{rr}=0$ at order $r^{-2}$ for $f_0(x,\theta)$.

\item[(d)] Solve the $d  r$ component of equation (\ref{cons}) at order $r^{-2}$ 
  for $k_0(x,\theta)$.  (Note that the differential equations for 
 $k_0(x,\theta)$ that arise in the $d  x$ and $d  \theta$ components of 
equation (\ref{cons}) at the preceding 
order $r^{-1}$ are now automatically satisfied.)

\end{itemize}

One then proceeds by iterating steps (a)--(d) at the next order, solving
algebraically for $g_0(x,\theta)$, $h_0(x,\theta)$, $f_1(x,\theta)$,
and $k_1(x,\theta)$, and so on ad infinitum.  The results for the first few
expansion coefficients are:
\bea
\bar f_1 &=& Gi + \log\tan\ft12\theta\,,\qquad
\bar g_1= \fft{\sin\theta}{\sqrt{G}}\,,\qquad
\bar h_1= \fft{\sqrt{G}\, \Gj\, \cos\theta -1}{A\,\sqrt{G}}\,,\nn\\
f_0&=& - \fft{[\sqrt{G}\, \Gj\, \cos\theta -1]^2}{2 A\, \sqrt{G}\,
\sin\theta}\,,\qquad
k_0 = \fft{\sqrt{G}\, [1- G \Gj^2\, \cos^2\theta]}{2A\, \sin\theta}
\,,\nn\\
g_0&=&\fft{[2+\sqrt{G}\, G'\, \Gj\, \cos^2\theta]\, \Gj}{
4 A\, \sqrt{G(x)}}\,,\nn\\
h_0&=& \fft{[1-\sqrt{G}\, \Gj\, \cos\theta][ 2\cos\theta +
\sqrt{G}\, G'\, \Gj\, \cos\theta + 
2\sqrt{G}\, \Gj\, \sin^2\theta]}{4 A^2\, G\, \sin\theta}\,,\nn\\
f_1&=& \fft{[1-\sqrt{G}\, \Gj\, \cos\theta]^2\, [4\sqrt{G}\, \Gj - G' +
2\sqrt{G}\, G'\, \Gj\, \cos\theta + G\, G'\, \Gj^2\,\cos^2\theta]}{
16 A^2\, G\, \sin^2\theta}\,, \nn\\
k_1&=& \fft{[G\, \Gj^2\, \cos^2\theta-1][4 \sqrt{G}\, \Gj -G' +
3G\, G'\, \Gj^2\, \cos^2\theta]}{16 A^2\, \sin^2\theta}\,,\nn\\
g_1&=&\fft{[2\sqrt{G}\, \Gj + G']^2 - 
   2 [G\, \Gj^2\, \cos^2\theta-1]^2\,G\, G''}{32 A^2\,G^{3/2}\, \sin\theta}\,,
\eea
where the arguments of $G$, $G'$, $G''$, $\Gj$ and $Gi$ are all $x$, 
defined implicitly in terms of $u$ and $\theta$ by equation (\ref{xdef}).  We
have obtained explicit results also for 
$(h_1,f_2,k_2,g_2,h_2,f_3,k_3,g_3,h_3)$.

The first few terms in the expansions (\ref{metcomp}) are given by
\begin{align}
C_{0}^\theta &=\fft{[2 G' + \sqrt{G}\, {G'}^2\, \Gj - 2 G^{3/2}\, G''\,
  \Gj]\, \cos\theta}{8 A\, \sqrt{G}\, \sin^2\theta}\,,\qquad
C_0^\phi =0\,,\nn\\
C_1^\theta&= \fft{{G'}^2\, \cos\theta}{16 A^2\, G\, \sin^3\theta} +
\fft{\Gj\, [9 {G'}^2 - 6 G'\,(2+3 G G'') + 8 G^2\, G''']\,\cos\theta}{
96 A^2\, \sqrt{G}\, \sin^3\theta} \nn\\
&+ \fft{\Gj^2\, (3 {G'}^2 -6G G'' -8)\,\cos\theta}{16 A^2\, \sin^3\theta}
+ \fft{\Gj^3\, G^{5/2}\,G'''\, \cos^3\theta}{12 A^2\, \sin^3\theta}\,.\nn\\
C_1^\phi &=0\,,\nn\\
C_{\theta\theta}&= -\fft{2\sqrt{G}\, \Gj + G'}{2 A\, \sqrt{G}\, \sin\theta}\,,
\qquad
C_{\theta\phi}=0\,,\qquad 
C_{\phi\phi}= 
 \fft{2\sqrt{G}\, \Gj + G'}{2 A\, \sqrt{G}}\sin\theta \,,\label{CCF}\\
D^{(1)}_{\theta\theta}&= -\fft{\Gj^4\, G^{5/2}\, G'''\, \cos^4\theta}{
48 A^3\, \sin^3\theta} - \fft{\Gj^3}{8 A^3\, \sin^3\theta} 
+\fft{\Gj^2\, (2 G^2\, G'''\, \cos^2\theta -9 G')}{48  A^3\, 
  \sqrt{G}\, \sin^3\theta} \nn\\
& - \fft{3\Gj\, {G'}^2}{32 A^3\, G\, \sin^3\theta}
-\fft{3{G'}^3 + 4 G^2\, G'''}{192 A^3\, G^{3/2}\, \sin^3\theta}\,,\nn\\
D^{(1)}_{\phi\phi} &= -D^{(1)}_{\theta\theta}\, \sin^2\theta\,,
 \qquad D^{(1)}_{\theta\phi}=0\,,\nn\\
F_0 = &\fft{12 G^3\, G'''\, \Gj^2\, \cos^2\theta + 6\sqrt{G}\, 
   [4-{G'}^2 + 2 G\, G''] \,\Gj +
  6G'\, (2+G\, G'')-3 {G'}^3-4 G^2\, G'''}{48 A\, \sqrt{G}\, \sin^3\theta}
\,.\nn
\end{align}

\subsection{Small mass expansion} \label{app:smallm}

It is hard to gain much insight from these expressions as they stand, since
$x$ is defined implicitly in terms of $u$ and $\theta$ by equation (\ref{xdef}).
One thing we can do is to consider an expansion in powers of the mass
parameter $m$ (or, to be precise, in powers of the small dimensionless 
quantity $m A$).  Expanding the integrand $G^{-3/2}$ in equation (\ref{Gjdef}) 
in powers of $m$ and then integrating term by term, we then have
\be
\Gj(x) = \fft{x}{\sqrt{1-x^2}} + \fft{m A\, (3x^2-2)}{(1-x^2)^{3/2}} +
{\cal O}(m^2)\,.\label{Gjm}
\ee
Conversely, we can then express $x$ perturbatively in $m$, in 
terms of $u$ and $\theta$, finding
\be
x = \fft{u A}{\sqrt{u^2 A^2 + \sin^2\theta}} +
       \fft{m A (u^2 A^2 -2 \sin^2\theta)}{u^2 A^2 + \sin^2\theta}
+{\cal O}(m^2)\,.\label{xm}
\ee
It is now straightforward to expand the expressions in (\ref{CCF}) 
for $C_0^I$, $C^{IJ}$ and $F_0$ in powers of $m$.  In particular, we
find
\be
C_0^\theta = -\fft{m\, \sin\theta\,\cos\theta}{
            (u^2A^2 + \sin^2\theta)^{3/2}} + {\cal O}(m^2)\,.
\ee
The expansion coefficient $F_0$, which is
related to the Bondi mass aspect, is given by
\be
F_0= \fft{m\, [(3 u^2 A^2+1)\, \sin^2\theta - 2 u^2 A^2]}{
     (u^2 A^2 + \sin^2\theta)^{5/2}} + {\cal O}(m^2)\,.
\ee
We also have
\be
C_{\theta\theta}= \fft{2m(2u^2 A^2 + \sin^2\theta)}{
\sqrt{u^2 A^2 + \sin^2\theta}\, \sin^2\theta} + {\cal O}(m^2)\,,\qquad
C_{\phi\phi}= -C_{\theta\theta}\, \sin^2\theta\,.
\ee

\section{Taub-NUT solution and the dyon solution}

The Taub-NUT metric was written as an expansion in Bondi coordinates in
\cite{godgodpoptn1}.  Introducing a complex null tetrad $(\ell,n,m,\bar m)$ 
for the Bondi metric (\ref{met}) in the form given in 
\eqref{NPframe}, and calculating the Weyl scalars in this frame at the leading couple of
orders in the $1/r$ expansion we find
\bea
\Psi_0 &=& -\fft{12\ell^2\, (\ell-i m)\,\tan^2\ft{\theta}{2}}{r^5}
+ \fft{60 i\, \ell^3\, (\ell -i m)\, \tan^2\ft{\theta}{2}}{r^6}
+\cdots
\,\nn\\
\Psi_1 &=& \fft{3(1+i)\ell\, (\ell-i m)\, \tan\ft{\theta}{2}}{r^4}
+\fft{12(1-i) \ell^2\, (\ell-i m)\tan\ft{\theta}{2}}{r^5}
+\cdots
\nn\\
\Psi_2 &=& -\fft{i\,(\ell-i m)}{r^3} - \fft{3\ell\, (\ell-i m)}{r^4} +\cdots
\qquad
\Psi_3  = - \fft{ 3(1-i)\, \ell\, (\ell- i m)\, \tan\ft{\theta}{2}}{2r^4}
  +\cdots
\nn\\
\Psi_4 &=& -\fft{3\ell^2\, (\ell-i m)\, \tan^2\ft{\theta}{2}}{r^5}+\cdots\,.
\label{Psitn}
\eea

  In the original Taub-NUT metric
\be
ds^2 = -h(\bar r)\, (d\bar t- 4\ell\, \sin^2\ft{\bar\theta}{2}\,
d\bar\phi)^2 + h(\bar r)^{-1}\, d\bar r^2 +
(\bar r^2+\ell^2)\, (d\bar\theta^2 + \sin^2\bar\theta\,d\bar\phi^2)\,
\label{tnmet0}
\ee
with\footnote{The tetrad vectors $\ell$ and $m$ should not be confused with
the NUT parameter $\ell$ and the mass $m$.  It should be clear from context
which is intended.}
\be
h(\bar r) = \fft{r^2-2m r -\ell^2}{\bar r^2+\ell^2}\,,
\ee
the single copy field strength $\cF=d\cA$ with
\be
\cA= -\fft{P\, (2\ell)^{-1}\, (\bar r^2-\ell^2) + Q \bar r}{\bar r^2 + \ell^2}\,
(d\bar t - 4 \ell\, \sin^2\ft{\bar\theta}{2}\, d\bar\phi)\label{Atn0}
\ee
satisfies the Maxwell equations $\nabla_\mu \cF^{\mu\nu}=0$ and describes
a configuration with electric charge $Q$ and magnetic 
charge $P$.\footnote{In fact
if the metric function $h(\bar r)$ in (\ref{tnmet0}) is modified to
$h= (\bar r^2 -2m \bar r -\ell^2 +P^2+Q^2)(\bar r^2+\ell^2)^{-1}$, the
metric and electromagnetic field also satisfy the coupled 
Einstein-Maxwell equations.}  Re-expressing the potential in the Bondi
coordinate system by employing the expansions given in \cite{godgodpoptn1},
\bea
\bar t &=& u + r + 2m\log r + 
\fft{\ell^2\, (4+3\cos\theta)\sec^4\ft{\theta}{2} -8m^2 -11\ell^2}{2r}+
\cdots\,,\nn\\
\bar\phi&=& \phi + \fft{\ell \sec^2\ft{\theta}{2}}{r} +\cdots\,,\nn\\
\bar r &=& r + \fft{\ell^2\, (3\cos\theta+5)\sin^2\ft{\theta}{2} \,
\sec^4\ft{\theta}{2}}{4r}+\cdots\,,\nn\\
\bar\theta &=& \theta -\fft{\ell^2\, \sin\ft{\theta}{2}\, 
 \sec^3\ft{\theta}{2}}{r^2} +\cdots\,,
\eea
we find, after making a gauge transformation that sets the $r$-component of
$\cA_\mu$ to zero~\footnote{We have calculated the components of the gauge field to higher order than the terms presented here.}, 
\be
\cA= \cA_\0 + \fft{\cA_\1}{r} + \fft{\cA_\2}{r^2}+\cdots\,,
\ee
with
\bea
\cA_\0 &=& 2P\, \sin^2\ft{\theta}{2}\, d\phi\,,\nn\\
\cA_\1&=& -Q\, du + 4\ell\, Q\, \sin^2\ft{\theta}{2}\, d\phi 
  -2\ell\, P\, \tan\ft{\theta}{2}\, d\theta\,,\nn\\
\cA_\2&=& \ell\, P\, du 
- 2\ell^2\, P\, (2+\cos\theta)\, \tan^2\ft{\theta}{2}\, d\phi
 -\ell^2\, Q\, (1+2\cos\theta)\, \sec^2\ft{\theta}{2}\, \tan\ft{\theta}{2}\,
d\theta\,.
\eea
The Maxwell scalars turn out to be given by
\bea
\Phi_0 &=& \fft{(1+i) \ell (P+i\, Q)\tan\ft{\theta}{2}}{r^3} 
 + \fft{3(1-i)\ell^2\, (P+i\, Q)\, \tan\ft{\theta}{2}}{r^4} +
{\cal O}(r^{-5})\,,\nn\\
\Phi_1 &=& -\fft{i\, (P+i\, Q)}{2r^2} - 
         \fft{\ell\, (P+i\, Q)}{r^3} +{\cal O}(r^{-4})\nn\\
\Phi_2 &=& -\fft{(1-i)\,\ell\, (P+i\, Q)\tan\ft{\theta}{2}}{2r^3} +
    {\cal O}(r^{-4}) \,.\label{maxtn}
\eea

We can see from the results for the Weyl scalars in (\ref{Psitn}) and
the Maxwell scalars in (\ref{maxtn}) that a relation of the form seen
in (\ref{PsiPhisqrel}) holds here also.  If we define
\be
R_0= \fft{\Phi_0^2}{\Psi_0}\,,\quad
R_1= \fft{\Phi_0\, \Phi_1}{\Psi_1}\,,\quad
R_2=\fft{\Phi_0\, \Phi_2 + 2 \Phi_1^2}{3\Psi_2}\,,\quad
R_3=\fft{\Phi_1\,\Phi_2}{\Psi_3}\,,\quad
R_4= \fft{\Phi_2^2}{\Psi_4}\,,
\ee
then to leading order in $1/r$ these are all the same:
\be \label{ScTN}
R_a =\frac{S}{3c}= \fft{(P+i\, Q)^2}{6(m+i\ell) r} + {\cal O}(r^{-2})\,,
\qquad\hbox{for all } a\,.
\ee

From 
\be
\Phi^{AB}\, \Phi_{AB} = \ft14 (F^{ab}\, F_{ab} +
                        {i\, ^*\!F}^{ab}\, F_{ab})= 
  \fft{(P+ i\, Q)^2}{2 r^4} - \fft{2 i\, \ell\, (P+i\, Q)^2}{r^5} +
  \cdots\,.
\ee
the Weyl double copy gives the scalar
\be
S = \fft{(P+i\, Q)^2}{2}\,\psi\,,\qquad 
     \psi=\fft1{r} -\fft{i\, \ell}{r^2}+\cdots \,,
\ee
which obeys $\square\psi=0$ in the flat background.  Comparing this expression with \eqref{ScTN}, we find agreement with
\begin{equation}
 c=m+i \ell.
\end{equation}

   It is interesting to note that in the expressions for $\Psi_0$,
$\Psi_1$, $\Psi_2$, $\Phi_0$ and $\Phi_1$, the first subleading order terms, 
along with the leading order terms, can be nicely combined by 
writing $r =\tilde r -i\, \ell$ and then expanding in inverse powers of 
$\tilde r$.  For example, the expansion for $\Psi_0$ in (\ref{Psitn}) 
becomes
\be
\Psi_0 = -\fft{12\ell^2\, (\ell-i m)\,\tan^2\ft{\theta}{2}}{\tilde r^5}
+{\cal O}(\tilde r^{-7})\,,
\ee
with analogous results in the other cases. 

\section{Rotating STU supergravity black holes} \label{app:STU}

In this appendix, we derive the Petrov type of the rotating STU black holes considered in section \ref{sec:STU}, as well as constructing the Bondi coordinates for the general solution.

\subsection{Petrov type of solution} \label{app:STUpet}

The Kerr-Newman black hole is type D, which means that when $s_1=s_2$ (corresponding to Kerr-Newman) the solution
becomes Petrov type D.  However, in general the rotating STU black holes are algebraically general (Petrov type I) 
as we show in this section.

  We use a null tetrad $(\ell,n, m,\bar m)$ given by the 
vectors\footnote{Note that here we are using a null tetrad adapted to the 
principal null directions of the metric, in which in particular
$\Psi_0=\Psi_4=0$.  This is not the same as the null tetrad in the Bondi 
frame that we are using in the bulk of the paper.}
\bea
\ell &=& \fft{r_1 r_2 + a^2}{\Delta}\, \del_t + \del_r + 
          \fft{a}{\Delta}\,\del_\phi\,,\qquad 
 n = \fft{\Delta}{2W}\, \Big[ \fft{(r_1 r_2+a^2)}{\Delta}\,\del_t 
 -\del_r + \fft{a}{\Delta} \,\del_\phi\Big]\,,
\nn\\
m &=&\fft{i}{\sqrt{2W}}\, \Big[ a \sin\theta\, \del_t -i \del_\theta
+\fft{1}{\sin\theta}\,\del_\phi\Big]\,,
\eea
with $\bar m$ being the complex conjugate of $m$.  The Weyl scalars
then turn out to be given by
\bea
\Psi_0&=& 0\,,\qquad \Psi_4=0\,,\nn\\
&&\nn\\
\Psi_1&=& \fft{i a M^2\, (s_1^2-s_2^2)^2\,\sin\theta}{
                               \sqrt2\,  W^{\ft52}}\,,\qquad
\Psi_3 = -\fft{i a M^2\, (s_1^2-s_2^2)^2\, \Delta\, \sin\theta}{
                 2\sqrt2 \, W^{\ft72}}\,,\label{Psi13}
\eea
and 
\bea
\Psi_2 &=& -\fft{M[ r+i a u + (s_1^2+s_2^2)\, (r+i a u -M)]\Big(
 [r+i a u + M(s_1^2 + s_2^2)]^2 - M^2(s_1^2-s_2^2)^2\Big)}{W^3} \nn\\
&&+ \fft{2M^2 (s_1^2-s_2^2)^2\Big(2\Delta -3i a (M-r) u - a^2(1+2u^2)
\Big)}{3 W^3}\,,\label{Psi2}
\eea
where $u=\cos\theta$.

As is clear from the expressions above, when $s_1=s_2$ all Weyl scalars vanish except for $\Psi_2$, which means that the Petrov type is D.  However, in the general case $s_1\ne s_2$, the Weyl scalars $\Psi_1$ and $\Psi_3$ 
do not vanish and so the solution is then not of type D.  However this, 
of itself, does not necessarily mean that the solution is 
algebraically general when $s_1 \neq s_2$, since we have not yet 
ruled out the existence of another principal null direction (PND) that is repeated, which would mean that the solution would be algebraically 
special (type II).  

In order to show that the solution is actually algebraically general, we 
can look at the null frame invariant 
\begin{equation}
 I^3-27J^2\,,
\end{equation}
where the quadratic invariant $I$ and cubic invariant $J$ are given in 
general by
\bea
I&=& \Psi_0 \Psi_4 - 4 \Psi_1 \Psi_3 + 3\Psi_2^2\,,\qquad
J=
  \begin{vmatrix} \Psi_4 & \Psi_3 &\Psi_2\\
          \Psi_3 & \Psi_2 & \Psi_1\\
          \Psi_2 & \Psi_1 & \Psi_0 \end{vmatrix}\,.
\label{IJdef}
\eea
Given an arbitrary null frame, there are then the possibilities \cite{stephani}
\begin{equation*}
 I^3-27J^2  \begin{cases}
              = 0 & \implies\textup{algebraically special} \\
              \neq 0 & \implies\textup{algebraically general}
             \end{cases}.
\end{equation*}

In our case, where $\Psi_0=\Psi_4=0$, we have
\bea \label{STUinv}
I^3- 27 J^2 = 4 \Psi_1^2\, \Psi_3^2\, (9\Psi_2^2 - 16 \Psi_1\, \Psi_3)\,.
\eea
Substituting the expressions in \eqref{Psi13} and \eqref{Psi2} gives
\bea
I^3- 27 J^2= a^4\, M^{10}\, (s_1^2-s_2^2)^8\, X\,,
\eea
where $X$ is a rather complicated expression whose precise
form is unimportant to the discussion, and which can be easily obtained from \eqref{Psi13} and \eqref{Psi2}.  Thus, the solutions are 
algebraically general when $s_1\ne s_2$.  In the Kerr-Newman case
$s_1=s_2$, the invariant does vanish, as expected.

It can be seen from (\ref{Psi13}) and (\ref{Psi2}) that at leading
order in $1/r$ at large distances, 
\bea \label{STUWeylfalloffs}
\Psi_1\sim\fft1{r^5}\,,\qquad \Psi_2\sim \fft1{r^3}\,,\qquad 
\Psi_3\sim\fft1{r^5}\,.
\eea
In a generic smooth asymptotically-flat metric the Weyl scalar 
$\Psi_s$ would fall off according to the peeling theorem as 
$\Psi_s\sim r^{s-5}$.  Thus in the rotating STU 
black holes $\Psi_1$ falls of faster than the leading-order peeling by
one inverse power of $r$, and $\Psi_3$ falls off faster than 
peeling by three inverse powers of $r$.  $\Psi_2$, on the other hand, 
falls off at precisely the $r^{-3}$ rate of leading-order peeling.  
It follows, therefore, that the metrics are approaching Petrov type D
asymptotically at large $r$. 

\subsection{Bondi coordinates}\label{app:STUBon}

Here, we cast the  metric (\ref{2charges1}) into the Bondi form defined 
by the metric \eqref{met} with fall-offs \eqref{metcomp} and determinant 
condition \eqref{det}.
We follow the procedure employed in \cite{godgodpoptn1} for the Kerr
black hole, where the metric is first expressed in terms of
an asymptotically spherical coordinate system (rather than 
the usual asymptotically spheroidal system of Boyer-Lindquist coordinates)
by means of the coordinate transformation
\bea
\tilde r^2 \, \sin^2 \tilde\theta =(\bar r^2+a^2)\, \sin^2\bar\theta\,,
\qquad
\tilde r^2\, \cos^2\tilde\theta =\bar r^2\, \cos^2\bar\theta\,,
\eea
and then further transformations are made, order by order in an expansion
in inverse powers of $r$, to put the resulting metric into the form
of eqn (\ref{met}). To the first few orders, we find that the required
coordinate transformations are given by
\bea
\bar t &=& u+r + 2M_B \, \log r -\fft{M^2\, [8+12(s_1^2+s_2^2) + 3(s_1^4+s_2^4) +
  18 s_1^2\, s_2^2]}{2 r} +\cdots\nn\\
\bar\phi &=& \phi - \fft{a M_B}{r^2} + \fft{2 a M^2\,[2 + 3(s_1^2+s_2^2)+
  s_1^4+ s_2^4 + 4 s_1^2\, s_2^2]}{3 r^3} +\cdots\nn\nn\\
\tilde r &=& r - M (s_1^2+s_2^2) +\fft{M^2\,(s_1^2-s_2^2)^2}{2 r} -
\fft{M a^2\, \sin^2\theta}{2 r^2} +\cdots\,,\nn\\\
\tilde\theta &=& \theta +\fft{M a^2\, (s_1^2+s_2^2)\sin 2\theta}{2 r^3} +
   \fft{a^2\, M^2\, (7 s_1^2 + 7 s_2^2 + 22 s_1^2 \, s_2^2) \sin 2\theta}{
12 r^4} +\cdots\,,
\eea
where we have defined the Bondi mass 
\bea
M_B = M\, (1 + s_1^2 + s_2^2)\label{Bonmass}
\eea
of the black hole.
We refer the reader to \cite{godgodpoptn1} for more 
details of the nature of the required expansions, which were obtained
there in the case of the Kerr black hole.  

  The components $g_{\mu\nu}$ of the resulting Bondi metric for the
rotating black holes at the first few orders turn out to be
\bea
g_{uu} &=& -1 + \fft{2 M_B}{r} - 
 \fft{2 M^2\, (s_1^2\, c_1^2 + s_2^2 \, c_2^2)}{r^2} +
\fft{M_B\, [2M^2\, (s_1^2-s_2^2)^2 - a^2 - 3 a^2 \cos2\theta]}{2 r^3}
+\cdots\,,\nn\\
g_{ur} &=& -1 + \fft{M^2\, (s_1^2 -s_2^2)^2}{2 r^2} +\cdots\nn\\
g_{u\theta} &=& \fft{3 a^2\, M_B\, \sin 2\theta}{2 r^2} +
 \fft{a^2\, M^2\, [18 (s_1^2+s_2^2) + 23 (s_1^4+s_2^4) -10 s_1^2\, s_2^2]
\sin2\theta}{12 r^3} +\cdots\,,\nn\\
g_{u\phi} &=& -\fft{2 a M_B\, \sin^2\theta}{r} +
\fft{2 a M^2\, (s_1^2\, c_1^2 + s_2^2\, c_2^2)\sin^2\theta}{r^2} +
\cdots\,,\nn\\
g_{\theta\theta}&=& r^2 -\fft{a^2\, M_B\, \sin^2\theta}{r} +
\fft{a^2\, M^2\, [6(s_1^2+s_2^2) + 5 (s_1^4 + s_2^4) +
                 2 s_1^2\, s_2^2]\sin^2\theta}{6 r^2} +\cdots\,,\nn\\
g_{\theta\phi} &=& -\fft{5 a^3\, M_B\, \cos\theta\,\sin^3\theta}{2r^2} +
\fft{12 a^3\, M^2\, (s_1^2\, c_1^2 + 
       s_2^2\, c_2^2)\, \cos\theta\,\sin^3\theta}{5 r^3} +\cdots\,,\nn\\
g_{\phi\phi} &=& r^2\, \sin^2\theta + \fft{a^2\, M_B\,\sin^4\theta}{r}
+ \fft{a^2 \, M^2\, [6(s_1^2+s_2^2) + 5(s_1^4+s_2^4) + 
             2 s_1^2\, s_2^2]}{6 r^2} +\cdots\,,
\eea
along with $g_{rr}$, $g_{r\theta}$ and $g_{r\phi}$ vanishing.
(We are only presenting a few leading order terms here; we have actually 
calculated the metric components to order $1/r^{10}$.)

We find that the 
first few coefficients in the expansions (\ref{metcomp}) are given by
\bea
F_0&=& -2M_B\,,\qquad 
F_1= \ft12 M^2 (4s_1^2 + 4 s_2^2 + 5 s_1^4 + 5 s_2^4 - 2 s_1^2 s_2^2)
  \,,\nn\\
\beta_0&=& -\ft14 M^2 (s_1^2-s_2^2)^2\,, \qquad \beta_1=0 \,,\nn\\
C_0^\theta&=&0\,,\qquad C_0^\phi=0\,,\qquad 
C_1^\theta=0\,,\qquad C_1^\phi= 2 a M_B\,,\nn\\
C_2^\theta&=& -3 a^2 M_B\, \sin\theta\cos\theta\,,\qquad
C_2^\phi= -2 a M^2\, (s_1^2\, c_1^2 + s_2^2\, c_2^2)\,,\nn\\
C_{IJ}&=&0\,,\qquad D^{(1)}_{\theta\theta}=- a^2\,M_B\,
\sin^2\theta\,,\qquad D^{(1)}_{\phi\phi}=-D^{(1)}_{\theta\theta}\, \sin^2\theta\,,
\qquad D^{(1)}_{\theta\phi}=0\,,\nn\\
D^{(2)}_{\theta\theta}&=& \ft16 a^2\, M^2\,[5 s_1^2\, c_1^2 + 5 s_2^2 \,c_2^2 +
  (s_1^2 + s_2^2)^2] \,\sin^2\theta\,,\qquad 
D^{(2)}_{\phi\phi} =-D^{(2)}_{\theta\theta}\, \sin^2\theta\,,\nn\\
D^{(2)}_{\theta\phi} &=& -\ft52  a^3\, M_B\, \sin^2\theta\,.
\eea

\section{Not quite Weyl double copy for Kerr-Newman}
\label{app:wdcKN}

The Weyl double copy \cite{lumonioc} has been formulated for vacuum spacetimes only, but analogous relations --- which do not necessarily have a double copy interpretation -- for non-vacuum solutions have been considered, including for the Reissner-Nordstr\"om solution of the Einstein-Maxwell equations \cite{Alawadhi2020}.  
Given that the rotating STU supergravity black holes we are considering 
here are non-vacuum, and that for certain choices of parameters they 
simplify to (same-charge) dyonic Kerr-Newman black holes, we consider, 
in this appendix, what happens if we try to apply the Weyl double copy for the Kerr-Newman solution to the Einstein-Maxwell theory.

The Kerr-Newman solution is given by the metric
\bea
ds^2 &=& -\fft{\Delta}{\Sigma}\, (dt - a\, \sin^2\theta\,
d\phi)^2 + \Sigma\,  \Big( \fft{dr^2}{\Delta} + d\theta^2\Big) 
   + \fft{\sin^2\theta}{\Sigma}\, [a\, dt - (r^2 +  a^2)d\phi]^2\,,\label{KN}\\
\eea
with
\begin{equation}
 \Delta = r^2 - 2Mr + a^2 + M^2 (q^2 + p^2), \qquad \Sigma = r^2 + a^2 \cos^2 \theta.
\end{equation}
Note that we have rescaled the electric and magnetic charges with a factor of the mass 
\begin{equation}
 Q = M q, \qquad P = M p,
\end{equation}
so that setting $M=0$ corresponds to Minkowski spacetime 
(in a spheroidal coordinate system).  The corresponding gauge potential is
\begin{equation}
A= \fft{- q M r (dt-a\, \sin^2\theta\,  d\phi)+ p M [a\, dt - (r^2 + a^2) d\phi]\, \cos\theta}{\Sigma}.
\label{KN:A}
\end{equation}

The single copy is of the same form as the gauge potential of the solution:
\begin{equation}
\cA= \ \fft{- \tilde{Q} r (dt-a\, \sin^2\theta\,  d\phi)+ \tilde{P} [a\, dt - (r^2 + a^2) d\phi]\, \cos\theta}{\Sigma}\,,
\label{KN:single}
\end{equation}
and it solves the Maxwell equation on both the Kerr-Newman and the Minkowski backgrounds, because it does not depend on $M$.  

In a null frame adapted to the principal null directions, given by
\begin{gather}
 \ell = \left(\frac{r^2+a^2}{\Delta} \right) \frac{\partial}{\partial t} + \frac{\partial}{\partial r} + \frac{a}{\Delta } \frac{\partial}{\partial \phi}, \qquad
 n = \frac{1}{2 \Sigma} \left( (r^2+a^2) \frac{\partial}{\partial t} -\Delta \frac{\partial}{\partial r}+ a \frac{\partial}{\partial \phi} \right)  \nn \\
 m = \frac{i}{\sqrt{2} (r+i a \cos \theta)} \left( a \sin \theta \frac{\partial}{\partial t}-i \frac{\partial}{\partial \theta}+ \frac{1}{\sin \theta} \frac{\partial}{\partial \phi}\right),
\end{gather}
the only non-zero Maxwell scalar is
\begin{equation}
 {\Phi_1} = \frac{ \tilde{Q}-i \tilde{P}}{2 (r- i a \cos \theta)^2},
\end{equation}
while the only non-zero Weyl scalar is 
\begin{equation}
 \Psi_2 = - \frac{M \left[r+i a \cos \theta -M \left(q^2+p^2\right)\right]}{(r-i a \cos \theta)^3 (r+i a \cos \theta)}\,.
\end{equation}
Hence, as is well-known, this solution is of type D.
Choosing 
\begin{equation}
 c = - 6 M i\,  (\tilde{Q}- i \tilde{P})^{-3/2}\,,
\end{equation}
the equation \eqref{PsiPhisqrel} yields
\begin{equation}
 S= \frac{3 i \sqrt{\tilde{Q}-i \tilde{P}}\ (r+i a \cos \theta )}{ \left[r+i a \cos \theta -M \left(q^2+p^2\right) \right]  (r-i a \cos \theta )}\,.
 \label{eq:SKN}
\end{equation}

Naively, there appears to be no obstacle to thinking of this as a double copy interpretation of the Kerr-Newman solution --- or, at least, of the metric part, since the gauge field \eqref{KN:A} would need to be similarly interpreted.  However, we need to also consider whether the single copy $S$ satisfies the wave equation on Minkowski spacetime. It is here that the subtlety appears. Recall equation \eqref{eq:boxS}, which follows from the vacuum type D Weyl double copy: $S$ does not depend on the curvature-generating parameters but the box operator does, which gives rise to the non-trivial right-hand side. When considering the flat spacetime box operator, $\Box\to\Box_0$, then $S$ satisfies the flat spacetime wave equation, as the zeroth copy should. In the Kerr-Newman example  \eqref{eq:SKN}, however, the $M$ dependence of $S$ means that $\Box_0S\neq0$, so we would need to consider also $S\to S_0$, which does not follow the expectation of a double copy, since the zeroth copy (like the single copy) should live in flat spacetime. This breakdown is unsurprising. On the one hand, the Weyl double copy has so far only been presented for vacuum solutions. On the other hand, it is 
not known whether `pure' Einstein-Maxwell theory (as opposed to the theory with also a dilaton and an axion/B-field) admits a double copy construction. Asymptotically, however, the dependence of the Kerr-Newman metric on the electric and magnetic charges is sub-leading, so we still expect a leading-order asymptotic double copy to exist. The difficulty arises in sub-leading orders in the expansion away from null infinity. This example shows also that being algebraically special is not enough for a full double copy interpretation. 

Notice that the Kerr-Newman black hole has a perturbative interpretation in terms of scattering amplitudes in Einstein-Maxwell theory \cite{Moynihan:2019bor,Chung:2019yfs}. Whether or not the black hole can be understood as a double copy follows directly from whether or not the relevant Einstein-Maxwell scattering amplitudes can be understood as a double copy, which is an open question.


\bibliographystyle{utphys}
\bibliography{Cmet}
\end{document}